\documentclass[12pt,preprint,showpacs,showkeys,nofootinbib]{revtex4}

\usepackage{epsfig}

\def\D{{\rm D}}

\begin{document}

\title{Writhe of center vortices and topological charge -- an
explicit example}

\author{Falk Bruckmann\footnotemark[1] and Michael Engelhardt\footnotemark[2]}

\affiliation{\parbox{1cm}{ } \\
\footnotemark[1] Instituut-Lorentz for Theoretical Physics,
Universiteit Leiden, P.O. Box 9506, NL-2300 RA Leiden, The Netherlands
\footnotetext[1]{\tt bruckmann@lorentz.leidenuniv.nl}
\\ \\
\footnotemark[2] Institut f\"ur Theoretische Physik,
Universit\"at T\"ubingen, Auf der Morgenstelle 14, 72076 T\"ubingen, Germany
\footnotetext[2]{\tt engelm@tphys.physik.uni-tuebingen.de} \\ \\ }

\begin{abstract}
\noindent
The manner in which continuum center vortices generate topological charge
density is elucidated using an explicit example. The example vortex
world-surface contains one lone self-intersection point, which contributes
a quantum $1/2$ to the topological charge. On the other hand, the
surface in question is orientable and thus must carry global topological
charge zero due to general arguments. Therefore, there must be another
contribution, coming from vortex writhe. The latter is known for the lattice
analogue of the example vortex considered, where it is quite intuitive.
For the vortex in the continuum, including the limit of an infinitely thin
vortex, a careful analysis is performed and it is shown how the contribution
to the topological charge induced by writhe is distributed over the vortex
surface.
\end{abstract}



\maketitle

\section{Introduction}
\noindent
Chromomagnetic center vortex degrees of freedom \cite{spag} furnish the basis
for one of the most attractive paradigms of the strong
interaction vacuum, the center vortex picture. Using the model assumption
that the two-dimensional closed vortex world-surfaces can be regarded as
random surfaces in four-dimensional (Euclidean) space-time, the principal
phenomena characterizing low-energy strong interaction physics can be
successfully described, i.e., confinement, the spontaneous breaking of chiral
symmetry and the axial $U_A (1)$ anomaly \cite{selprep,preptop,csb}. This
picture was inspired and is corroborated by lattice experiments
\cite{mcg,forc,forclap,temp,greensrev} in which the vortex content of lattice
Yang-Mills configurations is extracted and subsequently used to assess the
significance of the vortex degrees of freedom for the phenomenology of the
strong interaction vacuum.

In particular, the gluonic topological charge entering the $U_A (1)$ anomaly
can be understood in terms of the topology of the vortex world-surfaces
\cite{cont,preptop,corn,hr,csb}. Topological charge associated with center
vortex world-surfaces is generated in two different ways, self-intersections
of the surfaces and writhe. In realistic random surface ensembles, the
latter is statistically by far the more important \cite{preptop,roman}.

\begin{figure}
\epsfig{file=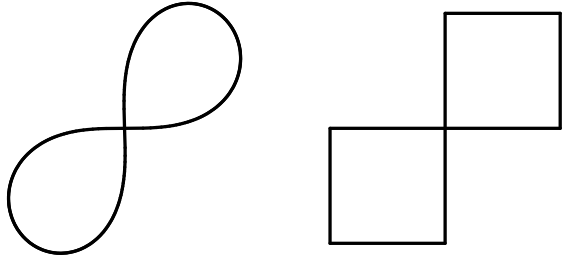,width=8cm}
\caption{Left: Continuum ``figure 8'' in a plane (e.g., a lemniscate);
right: coarse-grained lattice ``figure 8''.  \label{figintro} }
\end{figure}

On the other hand, the intuition gained hitherto from considering
lattice-generated ``cubistic'' vortex surfaces \cite{preptop,csb,hr} can be
misleading when trying to understand the more subtle writhe of continuum vortex
surfaces. To understand why, consider as a simple two-dimensional analogue the
continuum and lattice versions of a ``figure 8'', i.e. a closed 
self-intersecting line in a plane, as shown in Fig.~\ref{figintro}. As will be
discussed below, in four dimensions, lattice topological charge appears where
{\em the set of tangent vectors to the vortex spans all four dimensions of
space-time}. In the two-dimensional analogue of Fig.~\ref{figintro}, the
tangent vectors to the ``figure 8'' span both dimensions of the plane at the
self-intersection point, where two branches of the line provide two linear
independent tangent vectors. In addition, this is true for all the corners of
the lattice ``figure 8'', since the line bends in an abrupt way, resulting in
two linearly independent tangent vectors from one branch (as limits from
``before'' and ``after'' the point). Such a behaviour is the analogue of
writhe in the case of four-dimensional thin lattice vortices. However, at
first sight, it does not seem to be present for the continuum ``figure 8'',
where the tangent space is of course one-dimensional everywhere except at
the self-intersection point!

The purpose of the present note is to provide a simple, pedagogical,
explicit example of a continuum vortex configuration illustrating how
topological charge density arises. In particular, the contribution from
vortex writhe is elucidated. It turns out that continuum writhe is spread
smoothly over the surface and arises in connection with certain combinations
of {\em gradients of the tangent vectors}.

The example to be discussed here is formulated within the framework of 
$SU(2)$ gauge theory. The gluonic fields of this particular example will,
however, point into a fixed color direction throughout space-time. Thus,
the problem to be treated below is essentially an Abelian one.

\section{Description of the vortex world-surface}
\label{surfdesc}
\begin{figure}
\epsfig{file=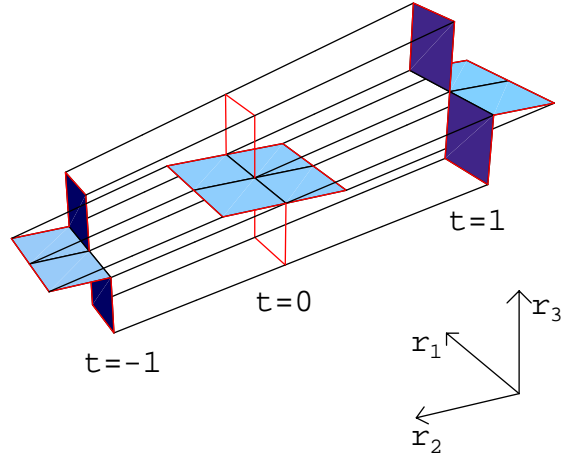,width=8cm}
\caption{Example vortex world-surface configuration composed of elementary
squares on a (Euclidean) four-dimensional hypercubic lattice, taken from
\cite{preptop}. At each lattice time slice, $t\in \{-1,0,1\} $, shaded
squares are part of the vortex surface. These
squares are furthermore connected to squares running in the time direction;
their location can be inferred most easily by keeping in mind that each
edge (lattice link) of the configuration is connected to exactly two
squares, i.e., the surface is closed. The surface possesses one isolated
point at which its self-intersects, at the center of the configuration,
and no self-intersection lines. Note that the two non-shaded squares at
$t=0$ are {\em not} part of the vortex; only the two sets of three links
bounding them are. These are slices at $t=0$ of surface segments running in
time direction from $t=-1$ through to $t=1$; sliced at $t=0$, these surface
segments show up as lines. \label{fig1} }
\end{figure}

\begin{figure}
\epsfig{file=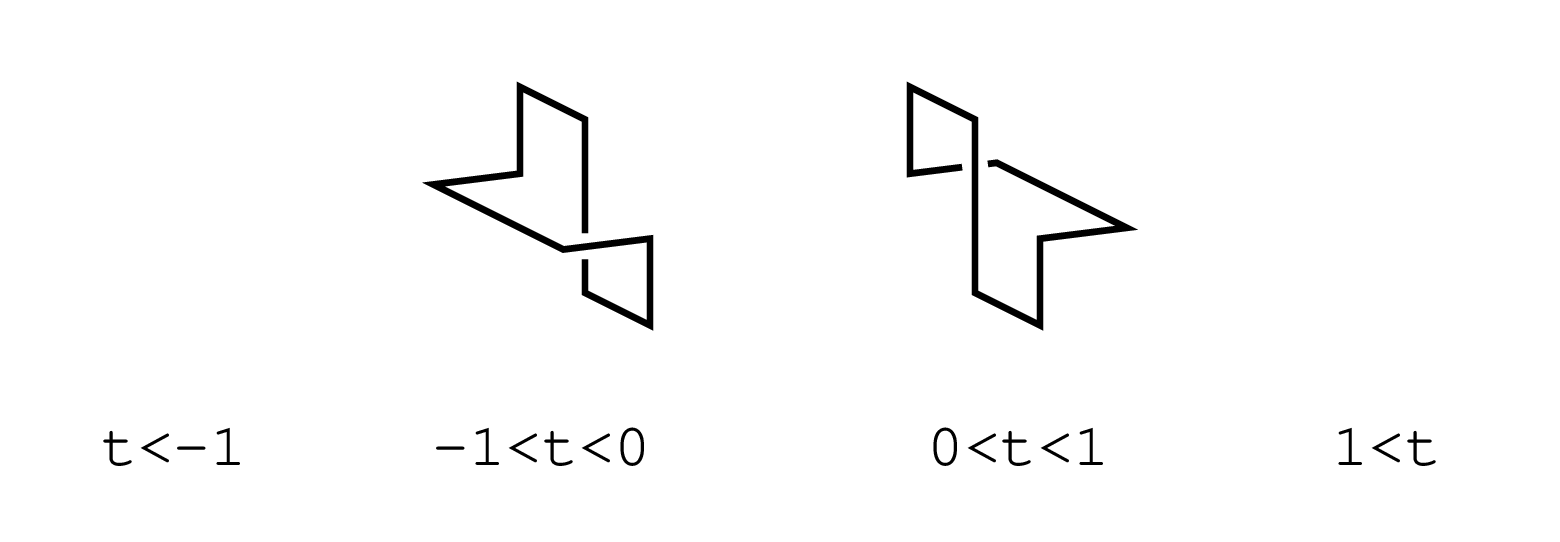,width=15cm}
\caption{Viewing the vortex world-surface depicted in Fig.~\ref{fig1}
in terms of the time evolution of a vortex line in three-dimensional
space, there are only four distinct stages of the time evolution,
during which the vortex shape remains constant. \label{fig2} }
\end{figure}

\noindent
Consider the vortex world-surface depicted in Fig.~\ref{fig1}, which was
first introduced in \cite{preptop} and subsequently also studied in
\cite{hr}. It is composed of elementary squares on a (Euclidean)
four-dimensional hypercubic lattice. Due to the coarse-grained nature of the
underlying lattice, changes in the vortex shape as time evolves can only
take place abruptly, at the times $t=-1, 0, 1$, cf.~Fig.~\ref{fig2}.

\begin{figure}
\epsfig{file=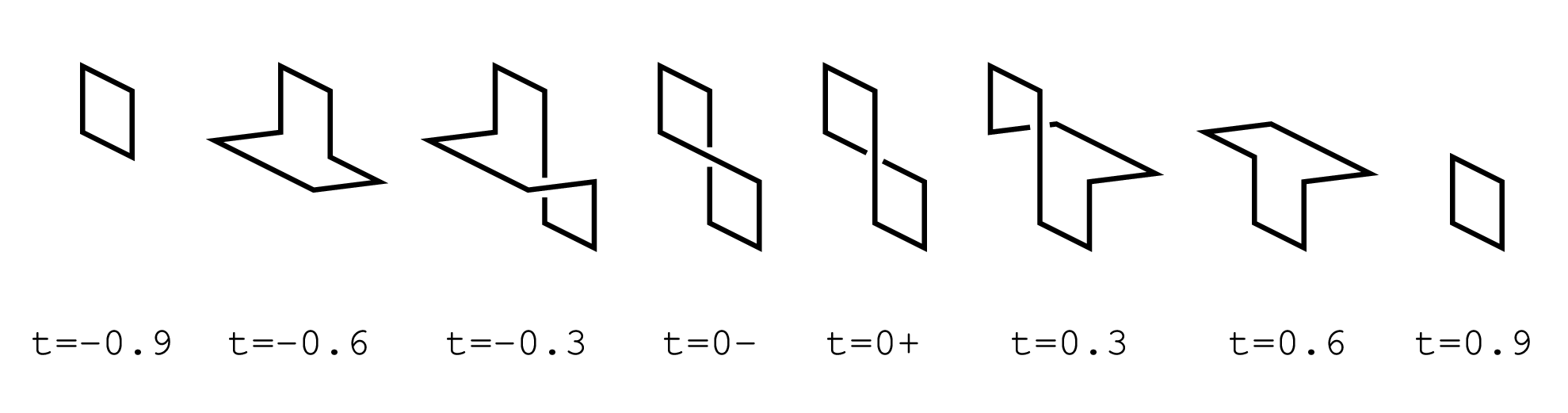,width=16cm}
\caption{Continuous, gradual time dependence of a vortex loop going through
the stages depicted in Fig.~\ref{fig2}, but without discontinuous
changes in the shape of the loop as time evolves. \label{fig3} }
\end{figure}

As a first step towards finding a continuum analogue, make the time
evolution gradual\footnote{A similar, but not identical, smoothing of
the time evolution was considered in \cite{hr}.}, as depicted in
Fig.~\ref{fig3}.

\begin{figure}
\epsfig{file=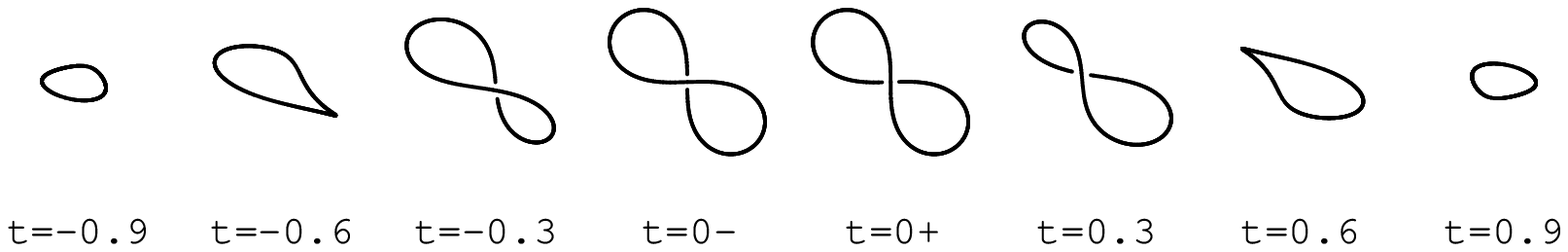,width=16cm}
\caption{Spatial shape of the loop displayed in Fig.~\ref{fig3} smoothed
out to yield an altogether smooth vortex world-surface. \label{fig4} }
\end{figure}

In other words, open a planar loop and then pull and simultaneously twist
the bottom of it in a corkscrew motion around a vertical axis by the
angle $\pi $, while holding the top fixed ($t=-0.9$ through $t=-0.3$).
Then let the vortex line intersect itself, as depicted by the images at
$t=-0.3$ through $t=0.3$. Afterwards, twist again by the angle $\pi $
such that the shapes at the times $t=0.3$ through $t=0.9$ are mirror images
of the shapes seen at the corresponding times $-t$, cf.~Fig.~\ref{fig3}.
Thus, one arrives again at a simple planar loop which then 
closes. Note that in the representation of Fig.~\ref{fig3}, and
also of Fig.~\ref{fig4}, in the last stage of the time evolution it is the
top of the loop which is corkscrewed back into the (approximately fixed)
bottom part. If one however considers the {\em relative} motion of the
bottom with respect to the top, one sees that the bottom rotates with
respect to the top by the angle $\pi $ {\em in the same direction} as
in the initial stage. The writhe in this configuration, to be discussed
in detail further below, comes from this $2\pi $ rotation of the two ends
of the loop with respect to each other; such a rotation of course is only
possible if one concomitantly allows the loop to self-intersect once.
As will be exhibited explicitly below, these two actions are each
associated with topological charge contributions of modulus $1/2$, which
cancel each other to give global topological charge zero.

As the second step towards a smooth continuum surface, all that remains
is to smoothen out the corners of the loops depicted in Fig.~\ref{fig3},
cf.~Fig.~\ref{fig4}.

This completes the construction of a smooth continuum analogue of the
lattice surface\footnote{Note that the discussion of topological charge
in \cite{cont} is in some respects too restricted in that it does not
encompass the smooth continuum surface discussed here, which contains
a single intersection point. In \cite{cont}, it is assumed that
intersection points of smooth surfaces occur in pairs. The formulation
in the published version of \cite{cont} is already sufficiently careful
to take into account the lattice version, Fig.~\ref{fig1}, by allowing
for other singular points on surfaces besides intersection points (by
that time, one of the authors had happened upon Fig.~\ref{fig1},
cf.~\cite{preptop}). Singular points are those points where the
tangent vectors to the vortex configuration span all four dimensions
of space-time.} shown in Fig.~\ref{fig1}. A convenient
parametrization of the complete world-surface is the following (the
source of this parametrization will be further commented upon below):
\begin{equation}
y (s_1 , s_2 ) = \sqrt{\sin 2s_2 } \left(
\begin{array}{c}
-\cos s_1 \ (\cos s_2 + \sin s_2 ) / \sqrt{2} \\
-\cos s_1 \ (\cos s_2 - \sin s_2 ) / \sqrt{2} \\
-\sin s_1 \ \cos s_2 \\
\sin s_1 \ \sin s_2
\end{array} \right) \ , 
\ \ \ s_1 \in [0,2\pi] \ , \ \ s_2 \in [0,\pi /2]
\label{paramet}
\end{equation}
In fact, Fig.~\ref{fig4} was generated using this parametrization; details
on how to extract closed vortex loops in three-dimensional space from
(\ref{paramet}) at fixed times $t\equiv y_1 (s_1 , s_2 )$ are given in
Appendix \ref{apploop}. In plotting Fig.~\ref{fig4}, the viewing angle of
the observer at the different times displayed was adjusted such as to best
match the viewing angle of Figs.~\ref{fig1}-\ref{fig3}.
The world-surface parametrized by eq.~(\ref{paramet})
will serve to illustrate how topological charge density arises from writhe
of smooth continuum vortex surfaces, which is the central objective of the
present work.

Specifying the space-time location of the vortex world-surface does not
completely determine a vortex configuration. It does determine the
structure of the moduli of the gluonic field strength tensor components,
$\mbox{Tr}_{color} \, F_{\mu \nu }^{2} $, for each $\mu , \nu $, which
are concentrated on the vortex surface in a way which will be described
more explicitly in the next section; however, it still leaves the color
direction of the field strength tensor components, $F_{\mu \nu }^{a} $,
free. Even if one assumes $F^a $ to be aligned with the three-direction in
color space\footnote{Since at generic points on a vortex surface, there is
only one nonvanishing field strength tensor component (details follow
further below), this can usually be achieved via a gauge transformation,
which will be assumed to have been done within the present treatment. The
field strength associated with the vortex world-surface (\ref{paramet})
constructed explicitly further below will indeed be of the form
$F^a = F\delta^{a3} $.}, $F^a = F\delta^{a3} $, then there still remains
a freedom in the sign of $F$, which can be viewed as a choice of the
{\em orientation} of the vortex world-surface.

Orientation can be assigned to a surface by locally associating a sense
of curl with the surface elements it is made up of. The reader is invited
to do so with the surface of Fig.~\ref{fig1}; for each elementary square
making up the surface, define a curl in the sense of a definite order of
running around the four edges of the square. If it is possible to define
curls for all squares such that all pairs of squares sharing an edge
(i.e.~neighbors) display matching orientation, then the surface is 
orientable. I.e., its global structure does not impose frustrations on the
attempt to align all neighboring surface elements, or in other words, the
attempt to {\em orient} the surface. This can be verified for the surface
shown in Fig.~\ref{fig1}, and it is equally true for the (topologically
equivalent) surface parametrized by eq.~(\ref{paramet}).

In terms of the associated vortex field strength, which will be constructed
explicitly in the next section, orienting the surface implies that the field
strength behaves in a continuous manner as one moves about the surface.
By contrast, lines on the surface at which the orientation flips would 
correspond to Dirac magnetic monopole singularities \cite{cont}. A
non-orientable surface structure actually forces the presence of monopoles
in the associated gauge field; however, also orientable surfaces can carry
monopoles (voluntarily, so to speak), if one chooses not to consistently
orient the surface but instead allows for lines on the surface at which
the orientation, i.e. the sign of the vortex field strength, flips.

The vortex surface parametrized by eq.~(\ref{paramet}) can also be arrived at
directly in the continuum, completely independent of the above connection to
a lattice surface. Namely, it results when casting a perturbed instanton into
the Laplacian Center Gauge, albeit with monopoles present on the surface
(which in fact is necessary in order to generate the unit topological charge
of the instanton). This is explained in more detail in Appendix \ref{pertinst}.
In the treatment below, on the other hand, no flips of orientation of the
field strength will be present on the (orientable) vortex surface parametrized
by eq.~(\ref{paramet}). The field strength will thus behave smoothly
as one moves about the surface. As a result of this simpler behavior
of the gluonic color vector, the configuration will carry {\em global }
topological charge zero. In fact, it is quite generally the case that
oriented vortex world-surfaces are associated with vanishing global
topological charge \cite{cont}. One way of seeing this is to recast the
global topological charge as an integral over the boundary of the
space-time manifold under consideration \cite{itzykson}. In
\cite{cont}, a construction of the vortex gauge field was given in which
this gauge field has support on a three-volume bounded by the vortex
surface, and with no Dirac string singularities in the gauge field 
{\em as long as the vortex surface is orientable}. Thus, for orientable
compact vortex surfaces (and correspondingly compact three-volumes spanning
them), the gauge field vanishes at the boundary of space-time and there
are no interior boundaries due to the absence of Dirac strings; therefore,
the global topological charge vanishes. By contrast, non-oriented vortex
surfaces force the presence of Dirac strings, thus providing the possibility
of nonvanishing topological charge via contributions from interior
integration boundaries.

\section{Field strength and topological charge}
\noindent
Chromomagnetic center vortices carry a field strength concentrated on the
vortex, where the nonvanishing field strength tensor component is the one
associated with the two space-time directions locally perpendicular to the
vortex surface. Given a surface parametrization $y(s)=y(s_1 , s_2)$ such
as (\ref{paramet}), a corresponding field strength can be constructed
explicitly as
\begin{equation}
F_{\mu \nu } (x) = \frac{\pi \sigma^{3} }{2}
\epsilon_{\mu \nu \kappa \lambda } \int d^2 s \,
\Sigma_{\kappa \lambda } (s) f(x-y(s))
\label{fsdef}
\end{equation}
with $d^2 s = ds_1 ds_2 $ and the (oriented) surface element
\begin{equation}
\Sigma_{\kappa \lambda } (s) \equiv \epsilon_{ab}
\frac{\partial y_{\kappa } }{\partial s_a }
\frac{\partial y_{\lambda } }{\partial s_b }
\label{sufel}
\end{equation}
where $\epsilon_{ab} $ represents the usual two-dimensional antisymmetric
symbol with $\epsilon_{12} =1$. Due to the combination of the two tangent
vectors to the surface, $\partial y / \partial s_1 $ and
$\partial y / \partial s_2 $ in (\ref{sufel}), with the $\epsilon $-symbol
in (\ref{fsdef}), the nonvanishing field strength tensor component is thus
indeed the one associated with the two space-time directions locally
perpendicular to the vortex surface. For instance, a surface locally running
in the 1-2 directions carries $\Sigma_{12}$ and therefore $F_{34}$.
On the other hand, $\sigma^{3} $ denotes the third Pauli matrix, encoding
the color structure of the vortex. Note thus that the field strength to be
used here points into a constant direction in color space,
$F^a = F\delta^{a3} $, rendering the present problem essentially
Abelian\footnote{In particular, also the gauge field generating the vortex
field strength (\ref{fsdef}) can be constructed to point into the
three-direction in color space \cite{cont}.}. Finally, the
{\em profile function} $f$ in (\ref{fsdef}) provides the freedom to
vary the transverse structure of the vortex field strength, since it
controls the value of the field strength at finite distances
$|x-y(s)|$ from the surface parametrized by $y(s)$.
To preserve the total flux carried by the vortex when $f$ is varied,
$f$ must be normalized,
\begin{equation}
\int d^4 z \, f(z) = 1 \ .
\end{equation}
To be definite, below a (four-dimensional) Gaussian profile function
\begin{equation}
f(z) = \frac{1}{a^4 \pi^{2} } e^{-z^2 /a^2 }
\label{profunc}
\end{equation}
will be used, with the variable $a$ controlling the thickness of the vortex;
in the limit $a\rightarrow 0$, (\ref{profunc}) reduces to a four-dimensional
delta function,
\begin{equation}
\lim_{a\rightarrow 0} f(z) = \delta^{4} (z) \ .
\label{thin}
\end{equation}
This corresponds to the limit of an infinitely thin vortex. For the moment,
the vortices will in fact be taken to be infinitely thin, even though
realistic, physically relevant vortices are not infinitely thin, but thick
in the sense that the field strength is smeared out to the vicinity of every
point $y(s)$. Further below, a thickening of the vortices will indeed play a
crucial role in resolving the subtleties involved in evaluating the
topological charge density arising from writhe of a thin vortex world-surface.

Two more remarks about the field strength (\ref{fsdef}) are in order before
proceeding. For one, a proper field strength must satisfy continuity of
flux, i.e., the Bianchi identity,
\begin{equation}
\epsilon_{\rho \tau \mu \nu } \partial_{\tau } F_{\mu \nu } = 0
\label{bianchi}
\end{equation}
where it has already been used that the problem is essentially an Abelian
one. The field
strength (\ref{fsdef}) satisfies (\ref{bianchi}), for any profile function 
$f$, as sketched in Appendix \ref{appbian}. Secondly, the normalization of
(\ref{fsdef}) has been adequately chosen such that the vortex indeed carries
flux corresponding to the center of the $SU(2)$ gauge group \cite{cont},
i.e., a Wilson loop circumscribing the vortex yields the phase $-1$. This
is demonstrated in Appendix \ref{appcent}.

The topological charge $Q$ and the corresponding topological charge density
$q(x)$ are defined as
\begin{equation}
Q = \frac{1}{32\pi^{2} } \int d^4 x \, \epsilon_{\mu \nu \kappa \lambda } \,
\mbox{Tr} \, F_{\mu \nu } F_{\kappa \lambda }
\equiv \int d^4 x \, q(x) \ .
\label{topq}
\end{equation}
In view of this expression, topological charge density is generated at those
points in space-time where there are nonvanishing field strength tensor
components $F_{\mu \nu } $, $F_{\kappa \lambda } $ such that the indices
$\{ \mu , \nu , \kappa , \lambda \} $ span all four space-time directions.
Translating this (naively) into the properties of (thin)
vortex surfaces, in view of the discussion so far, the same must hold
for the surface elements $\Sigma_{\mu \nu } $, $\Sigma_{\kappa \lambda }$ of
the vortex. In other words (in analogy to the two-dimensional toy model of
the introduction), topological charge is ostensibly generated at those points
in space-time where the set of tangent vectors to a thin vortex surface
configuration spans all space-time directions \cite{preptop}.

This way of stating the characteristics of a vortex surface necessary for
generating topological charge is quite adequate in some circumstances.
For example, as will be seen in the example discussed below, for lattice
surfaces it is entirely sufficient, and technically convenient, to consider
idealized, infinitely thin surfaces for the purpose of evaluating the
topological charge\footnote{Note also that the global topological charge can
be recast as an integral over the boundary of the space-time manifold under
consideration \cite{itzykson}; therefore, it should be independent of any
specific assumptions about the vortex transverse profile, which, after all,
merely corresponds to a local deformation of the gauge field in the vicinity
of the vortex.}, and to look for the points where the set of tangent vectors
spans all four space-time directions. Furthermore, even when considering
general continuum vortex surfaces, {\em intersection points} are
adequately described by this characterization. For instance, a surface
locally running into the 1-2 directions carries $F_{34} $; a surface
locally running into the 3-4 directions carries $F_{12} $. If the two
happen to intersect at a point, the product $F_{12} F_{34} $ is nonvanishing
and topological charge is generated. Such intersection points will be
present for lattice surfaces as well as thin and thick vortices in the
continuum (where the surfaces need not cross perpendicularly). They always
contribute $\pm 1/2 $ to the topological charge (even when the crossing is
not perpendicular \cite{cont}), where the sign just depends on the relative
orientation of the surfaces.

On the other hand, one must not be too naive in applying the above
characterization, in the case of general continuum surfaces, to other
contributions to the topological charge, which will be subsumed under
the term {\em writhing} contributions\footnote{Writhe is a property of loops
in three dimensions which enters the topological charge in a 3+1 dimensional
picture \cite{cont}.}. The corresponding topological charge density is rather
hidden in the singular nature of the field strength associated with a thin
continuum vortex surface, as will be discussed in detail further below. As
a result, it is easily missed if one naively just looks for those points
in the surface configuration where the set of tangent vectors spans all four
space-time directions. It is the main objective of this work to elucidate
the writhing contributions using the example surfaces which were introduced
in section \ref{surfdesc}.

\subsection{Lattice surface}
\label{latsec}
\noindent
As already mentioned above, finding the set of points where the tangent
vectors to the surface configuration span all four space-time directions
is sufficient to capture all contributions to the topological charge in the
case of lattice surfaces made up of elementary squares. In this case, the
topological charge density is concentrated at lattice sites; each instance
of two squares sharing a lattice site and their combined tangent vectors
spanning all four space-time directions contributes a ``quantum'' $\pm 1/32$
to the topological charge \cite{preptop}. As before, the sign depends on the
relative orientation of the squares.

\begin{figure}
\epsfig{file=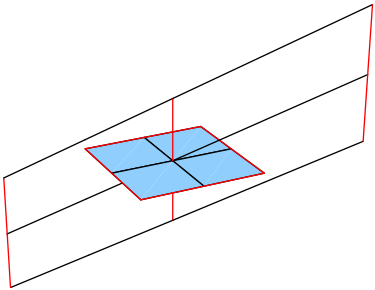,width=8cm}
\caption{Detailed view of the intersection point in
Fig.~\ref{fig1}. \label{fig5} }
\end{figure}

Considering the lattice configuration of Fig.~\ref{fig1}, one notices a
self-intersection point at the center of the configuration with $Q=1/2$.
This contribution results from four elementary squares connected to the
intersection point extending into the 1-2 directions and four elementary
squares connected to the intersection point extending into the
3-4 directions, cf.~the detailed view displayed in Fig.~\ref{fig5}.
This indeed amounts to 16 pairs of perpendicular squares, i.e., $Q=16/32=1/2$.

\begin{figure}
\epsfig{file=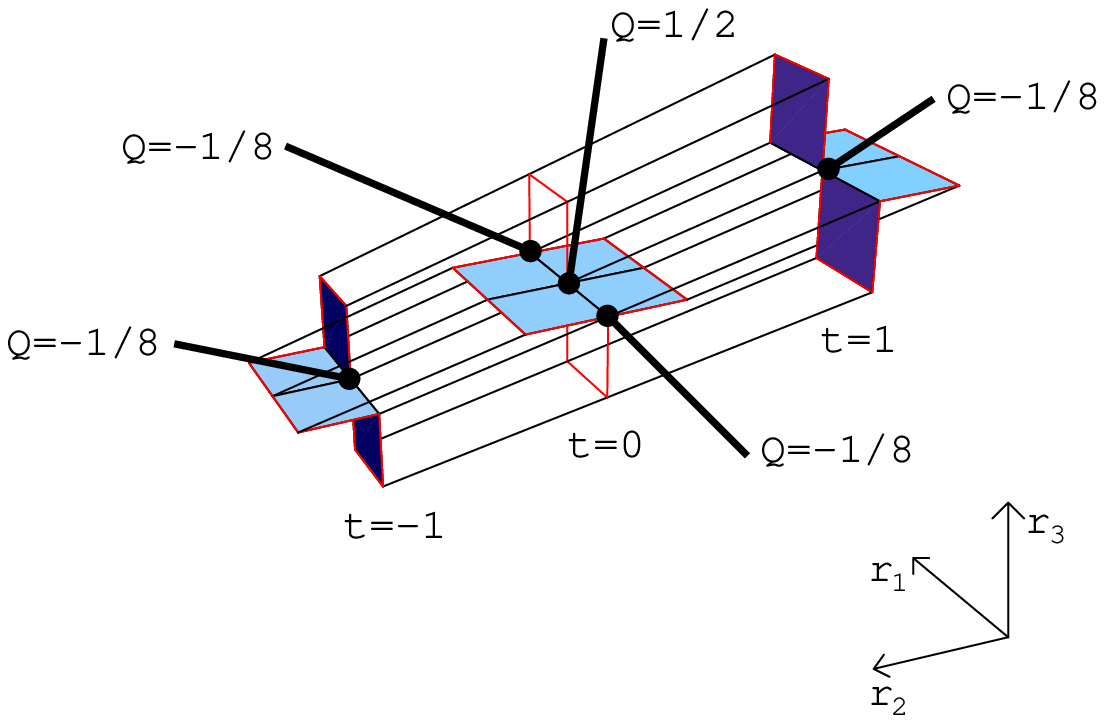,width=14cm}
\caption{Topological charge contributions from vortex self-intersection
and writhe present in the world-surface depicted in Fig.~\ref{fig1}.
To fully determine the signs of the contributions, the relative orientations
of the elementary squares making up the surface must be fixed; this was
done (starting from an initial square of arbitrary orientation) in such
a way as to completely orient the surface, i.e., there are no edges shared
by two squares at which the orientation flips. \label{fig6} }
\end{figure}

The complete set of topological charge contributions associated with the
surface depicted in Fig.~\ref{fig1} is shown in Fig.~\ref{fig6}.
Thus, apart from the intersection point at the center of the configuration,
there are four writhing contributions concentrated at other lattice sites.
At those sites, the vortex surface bends in such a way that it generates
the perpendicular squares needed for a topological charge contribution within
one single branch of the surface. A closer look reveals that there are four
pairs of perpendicular squares at each such point, i.e., topological charge
contributions $Q=4\times (-1/32)=-1/8$. Together, these writhing contributions
exactly cancel the one from the intersection point, such
that $Q_{global}=1/2-4/8=0$. This is in agreement with the general
statement that oriented vortex world-surfaces are associated with vanishing
global topological charge \cite{cont}.

Note thus that, on the lattice, writhe is concentrated at points on the
vortex surface. For smooth continuum surfaces, the topological charge
density associated with writhe will be spread out smoothly on the surfaces
corresponding to the more gradual way in which they twist; this will become
clear in the subsequent sections.

\subsection{Continuum surface}
\label{contsec}
\noindent
Having discussed the topological charge of the coarse-grained lattice
surface displayed in Fig.~\ref{fig1}, one may ask how this treatment
translates to the continuum analogue parametrized by eq.~(\ref{paramet}).
If one again starts from the statement that topological charge is generated
where the tangent vectors to the surface span all four space-time directions,
the contribution from the intersection point is as evident as before. However,
one may be (mis-) led to conclude that there are no further contributions;
after all, in the case of a smooth continuum surface, away from 
self-intersection points, there is always a well-defined unique
two-dimensional tangent plane at each point. Nowhere do the tangent vectors
span all four space-time directions. Where is the contribution from
vortex writhe\footnote{The authors acknowledge R.~Bertle for insisting
this question be answered, thus sparking the present investigation.}?

Clearly, it would be too naive to declare it absent; for one, the smooth
configuration shown in Fig.~\ref{fig4} is topologically equivalent to the
lattice version depicted in Fig.~\ref{fig1}. Also on general grounds, an
orientable surface such as the one of Fig.~\ref{fig4} must have global
topological charge zero \cite{cont}; therefore, there {\em must} be writhing
contributions canceling the contribution from the self-intersection point
which is certainly there. The point is that the writhing contributions are
rather hidden in the singular structure of the field strength corresponding
to an infinitely thin vortex surface, which has been implicitly assumed
in the above discussion.

To properly extract the writhing contributions,
one must view the infinitely thin vortex surfaces as an idealized limiting
case of general thick vortices (in fact, physically relevant vortices have
a finite thickness). When the vortex field strength is smeared out
transversally, there will of course in general be a certain overlap between
field strengths originating from the smearing of neighboring points on the
original thin surface. This leads to a nonvanishing topological density
$\epsilon_{\mu \nu \kappa \lambda }  F_{\mu \nu } F_{\kappa \lambda } $
even if the underlying thin surface is smooth, as long as it curves suitably,
i.e.~writhes, in the space-time region under consideration. If one now makes
the vortices thinner, then the overlap regions shrink, but at the same time,
the modulus of the field strength increases such that in the thin limit,
a finite topological density
remains on the thin vortex surface. This in fact must happen such that the
global topological charge is entirely independent of the vortex thickness,
including the infinitely thin limit -- after all, a topological quantity
should be independent of local deformations such as a thickening of the
vortex. The main purpose of the present treatment lies in demonstrating all
this explicitly using the specific example vortex world-surface introduced
in eq.~(\ref{paramet}). First, thick vortices will be treated, in which
the width of the profile functions $a$, cf.~eq.~(\ref{profunc}), can still be
varied. Then, the thin limit $a\rightarrow 0$ will be considered.

\subsubsection{Thick vortex}
\noindent
Inserting the field strength
(\ref{fsdef}) into the topological charge density (\ref{topq}), one obtains
\begin{equation}
q(x) = \frac{1}{16} \int d^2 s \int d^2 s^{\prime } \,
\epsilon_{\rho \tau \beta \gamma } \,
\Sigma_{\rho \tau } (s) \Sigma_{\beta \gamma } (s^{\prime } )
f(x-y(s) ) f(x-y(s^{\prime } ) )
\label{qdens}
\end{equation}
after having taken the color trace,
$\mbox{Tr} \, (\sigma^{3} \sigma^{3} ) =2$,
and having simplified the Lorentz structure with the help of
\begin{equation}
\epsilon_{\mu \nu \kappa \lambda } \epsilon_{\mu \nu \rho \tau }
\epsilon_{\kappa \lambda \beta \gamma } =
2(\delta_{\kappa \rho } \delta_{\lambda \tau } -
\delta_{\kappa \tau } \delta_{\lambda \rho } )
\epsilon_{\kappa \lambda \beta \gamma }
= 4 \epsilon_{\rho \tau \beta \gamma } \ .
\end{equation}
Appendix \ref{appess} contains the tedious, but straightforward, algebra
involved in inserting the specific parametrization (\ref{paramet}) into the
surface elements and subsequently reducing (\ref{qdens}) to
\begin{eqnarray}
q(x) &=& \frac{1}{4} \int d^2 s \int d^2 s^{\prime } \,
f(x-y(s) ) f(x-y(s^{\prime } ) ) \ \cdot \label{q3trig} \\
& & \left[
\sin 2s_2 \sin 2s_{2}^{\prime } \sin^{2} (s_1 -s_{1}^{\prime } )
-\sin 3(s_2 -s_{2}^{\prime } ) \sin (s_2 -s_{2}^{\prime } ) \right] \ .
\nonumber
\end{eqnarray}
To proceed from this simple form for the topological density,
explicit profile functions $f$ must be inserted. Using eq.~(\ref{profunc}),
one needs to evaluate the convolution of the Gaussian profile functions in
order to obtain the topological charge $\int d^4 x\, q(x)$,
\begin{eqnarray}
& & \int \! d^4 x \, f(x\! -\! y(s) ) f(x\! -\! y(s^{\prime } ) ) =
\label{convol} \\
& & \ \ \ \ \ = \frac{1}{4\pi^{2} a^4 }
\exp \left( -\frac{1}{a^2 } 
\left( \frac{1}{2} (y(s))^2 + \frac{1}{2} (y(s^{\prime } ))^2
-y(s) y(s^{\prime } ) \right) \right) \nonumber \\
& & \ \ \ \ \ = \frac{1}{4\pi^{2} a^4 } \exp \left( -\frac{1}{a^2 }
\left( \frac{1}{2} \sin 2s_2 + \frac{1}{2} \sin 2s_{2}^{\prime }
-\sqrt{\sin 2s_2 \sin 2s_{2}^{\prime } }
\cos (s_2 - s_{2}^{\prime } ) \cos (s_1 - s_{1}^{\prime } ) \right) \right)
\nonumber
\end{eqnarray}
where the parametrization (\ref{paramet}) has been put in. Inspecting
(\ref{q3trig}) and (\ref{convol}), one notices that the topological charge
is now an integral over $s,s^{\prime } $ in which the integrand depends only
on the difference $s_1 - s_{1}^{\prime } $, and not on the individual angles.
Therefore, by substituting $p = s_1 - s_{1}^{\prime } $ and
$q = s_1 + s_{1}^{\prime } $, with the Jacobian
$dp\, dq = 2 ds_1 \, ds_{1}^{\prime } $, one can carry out one of the
integrations via the general identity
\begin{equation}
\int_{0}^{2\pi } ds_1 \int_{0}^{2\pi } ds_{1}^{\prime } \,
g (s_1 - s_{1}^{\prime } ) = 
\frac{1}{2} \int_{-2\pi }^{2\pi } dp \, g(p)
\int_{|p|}^{4\pi - |p|} dq \, =
2\int_{0}^{2\pi } dp \, g(p) (2\pi -p)
\label{triangle}
\end{equation}
where in the last equality, the $p$-integration interval has been halved
using the fact that in the case considered here, $g(p) \equiv g(|p|)$.
Combining (\ref{q3trig}), (\ref{convol}) and (\ref{triangle}), the
topological charge reduces to
\begin{eqnarray}
Q\! &=& \! \frac{1}{8\pi^{2} a^4 } \!
\int_{0}^{\pi /2} \! ds_2 \! \int_{0}^{\pi /2} \! ds_{2}^{\prime } \!
\int_{0}^{2\pi } \! dp \,
\left[ \sin 2s_2 \sin 2s_{2}^{\prime } \sin^{2} p
-\sin 3(s_2 \! -\! s_{2}^{\prime } ) 
\sin (s_2 \! -\! s_{2}^{\prime } ) \right] \cdot
\label{qfinal} \\
& & \ \ \ \ \ (2\pi -p) \exp \left( -\frac{1}{a^2 }
\left( \frac{1}{2} \sin 2s_2 + \frac{1}{2} \sin 2s_{2}^{\prime }
-\sqrt{\sin 2s_2 \sin 2s_{2}^{\prime } }
\cos (s_2 - s_{2}^{\prime } ) \cos (p) \right) \right) \ .
\nonumber
\end{eqnarray}
This integral is easily evaluated numerically for diverse $a$, and always
vanishes, independent of the choice of $a$, as was to be expected from the
discussion of the lattice version of the vortex surface in the previous
section. Therefore, the integral must include writhing contributions
which serve to cancel the contribution of the self-intersection point at
the center of the surface configuration.

In order to exhibit the writhe in more detail, it is useful to evaluate the
topological charge density $q(x)$ numerically for a thick vortex profile
and visualize the result.
For this purpose, it is useful to rewrite eq.~(\ref{q3trig}). Working
with (\ref{q3trig}), one would need to evaluate a four-dimensional
integral for each space-time point $x$; however, the task can be greatly
simplified by decomposing the square brackets in (\ref{q3trig}) using
\begin{eqnarray}
\sin^{2} (s_1 -s_{1}^{\prime } ) &=& \frac{1}{2} \left(
1-\cos 2s_1 \cos 2s_{1}^{\prime } -\sin 2s_1 \sin 2s_{1}^{\prime } \right) \\
\sin 3(s_2 -s_{2}^{\prime } ) \sin (s_2 -s_{2}^{\prime } ) &=&
(\sin 3s_2 \cos 3s_{2}^{\prime } - \cos 3s_2 \sin 3s_{2}^{\prime } ) \,
(\sin s_2 \cos s_{2}^{\prime } - \cos s_2 \sin s_{2}^{\prime } ) \ .
\nonumber
\end{eqnarray}
By multiplying out all the terms which thus result in the square brackets
and letting the integrations act on them separately, the topological charge
density reduces to a sum over terms in which the four-dimensional integrals
factorize into pairs of two-dimensional integrals over the unprimed and the
primed variables, respectively.

\begin{figure}
\begin{center}
\epsfig{file=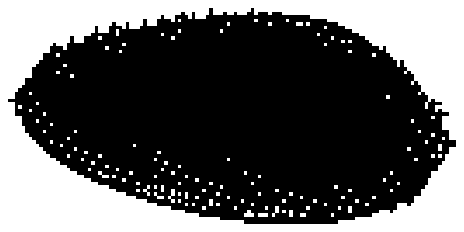,width=4cm}
\hspace{0.5cm}
\epsfig{file=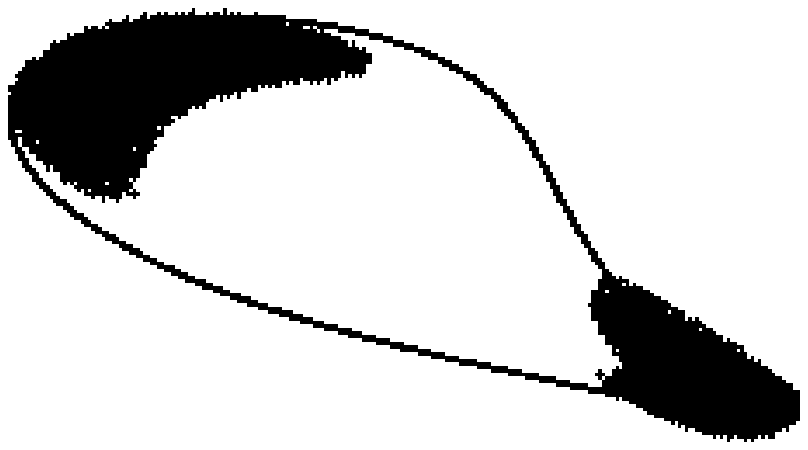,width=4cm}
\end{center}
\caption{Time slices of the topological charge density induced by the vortex
field strength (\ref{fsdef}) with the vortex world-surface parametrization
(\ref{paramet}) and a vortex thickness of $a=1/5$, cf.~(\ref{profunc}). At
the times $t=-0.9$ (left) and $t=-0.6$ (right), all points in space are
plotted at which the modulus of the topological charge density $|q(x)|$
exceeds unity. At the two times displayed, these are actually all points
with $q(x) < -1$; in the companion figure Fig.~\ref{fig8} below, depicting
further time slices, also points with $q(x) > 1$ will be seen. To guide the
eye, also the corresponding time slices of the thin vortex world-surface
parametrization are plotted, identical to the plots in Fig.~\ref{fig4} at
the corresponding times. Thus, the viewing angle of the observer matches
the one adopted in Fig.~\ref{fig4}. \label{fig7} }
\end{figure}

\begin{figure}
\begin{center}
\epsfig{file=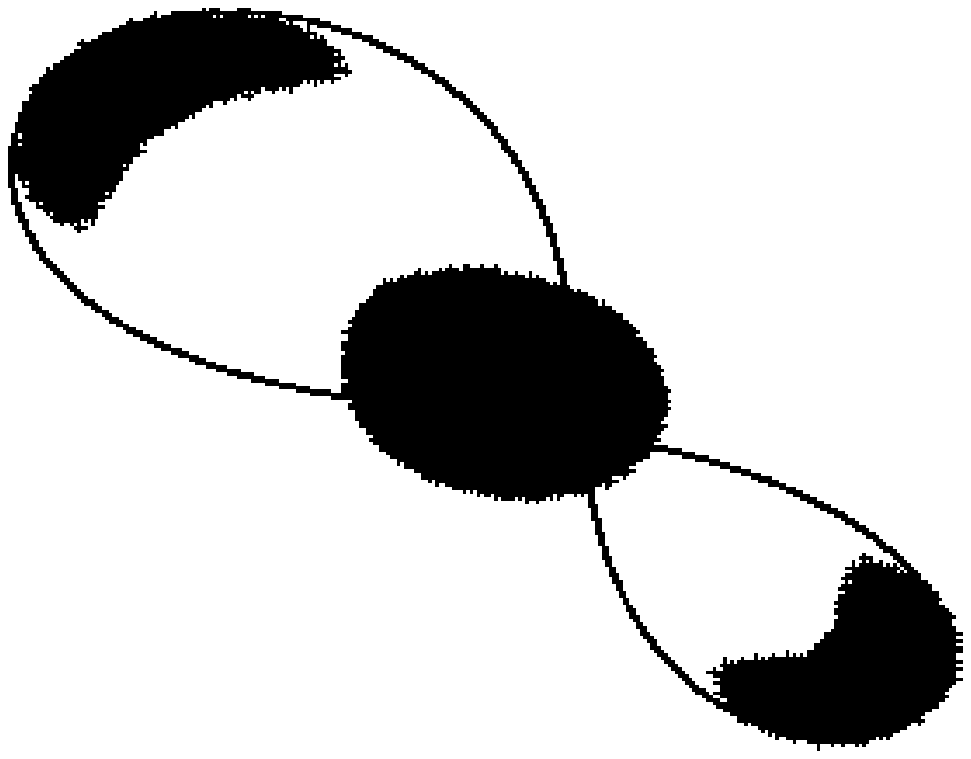,width=4cm}
\hspace{0.5cm}
\epsfig{file=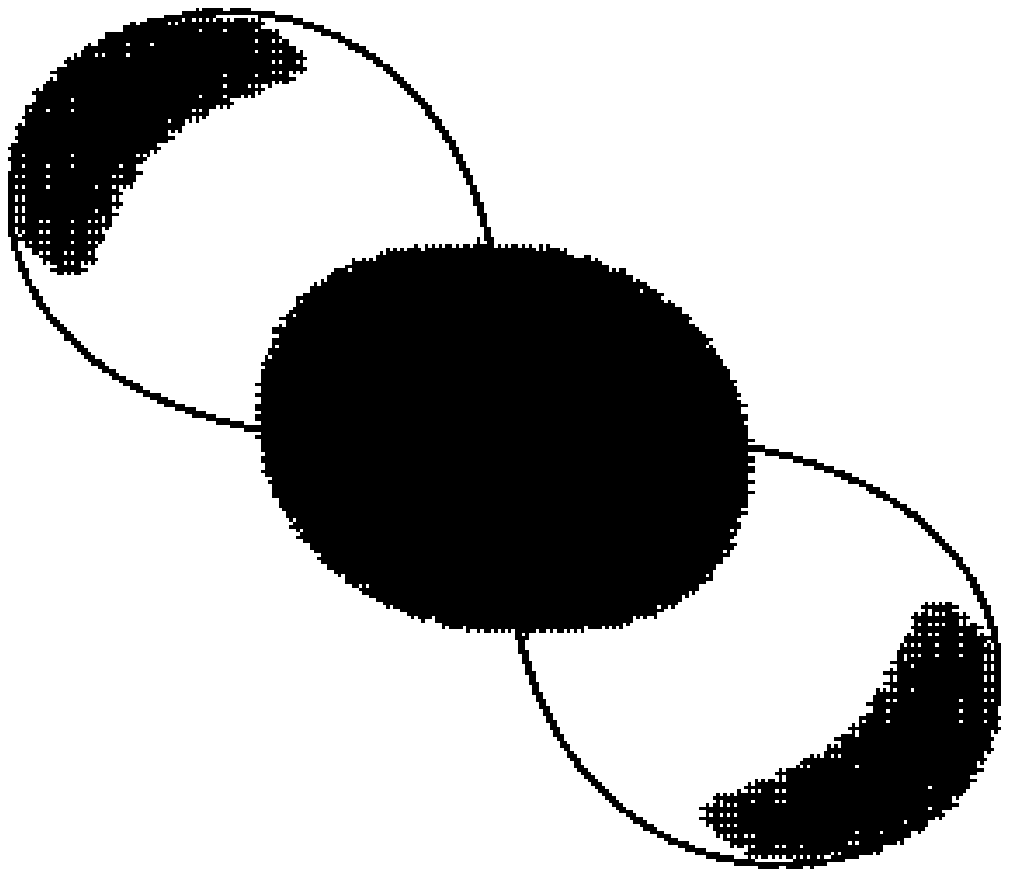,width=4cm}
\hspace{0.5cm}
\epsfig{file=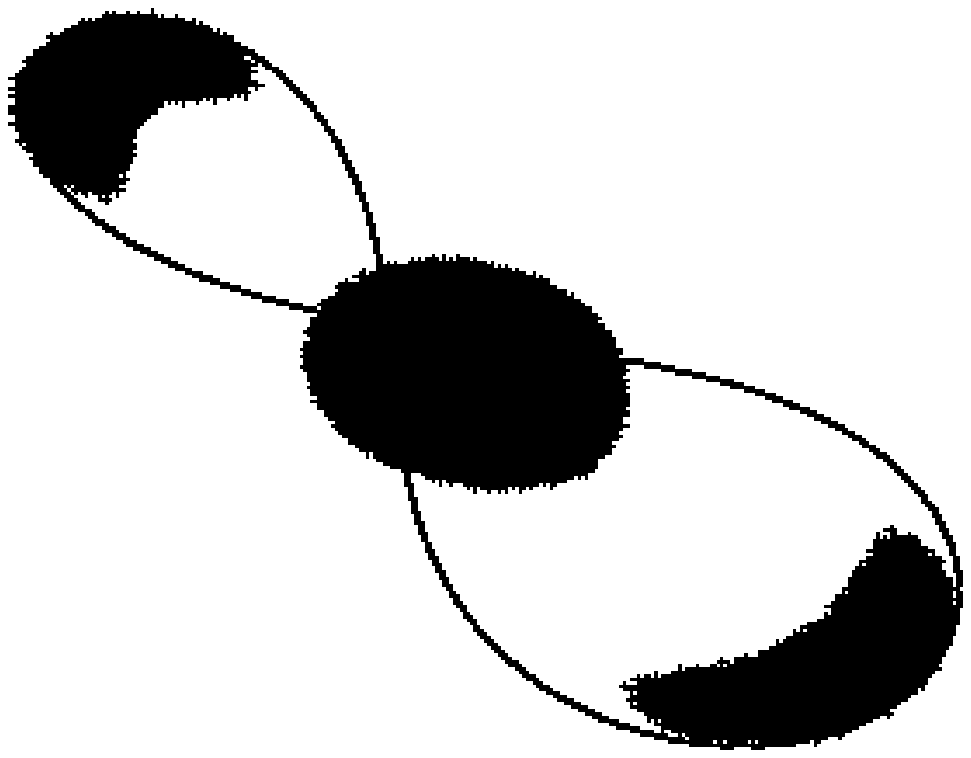,width=4cm}
\end{center}
\caption{As Fig.~\ref{fig7}, at the times $t=-0.3$ (left), $t=0$ (center)
and $t=0.3$ (right). At these times, both points with $q(x) > 1$ (at the
center of the configuration, induced by the vortex self-intersection), and
points with $q(x) < -1$ (kidney-shaped lumps at the periphery of the
configuration, induced by vortex writhe) are present. \label{fig8} }
\end{figure}

\begin{figure}
\begin{center}
\epsfig{file=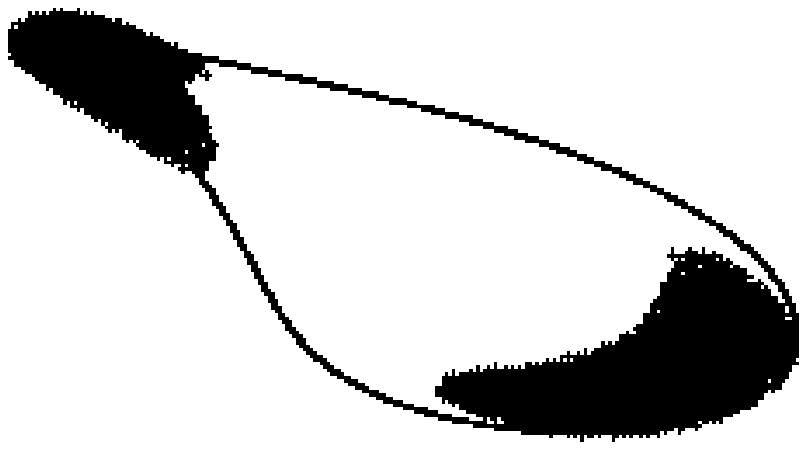,width=4cm}
\hspace{0.5cm}
\epsfig{file=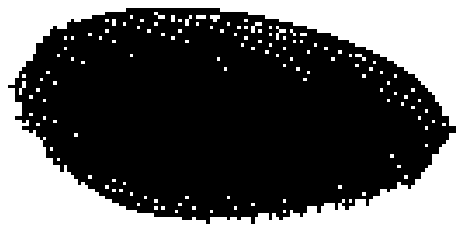,width=4cm}
\end{center}
\caption{As Fig.~\ref{fig7}, at the times $t=0.6$ (left) and $t=0.9$
(right). \label{fig9} }
\end{figure}

Using this numerically convenient form, the topological charge density was
evaluated for a grid of points in space at selected times $t\equiv x_1 $,
where, to be definite, the value $a=1/5$ was used in the profile functions
$f$, cf.~(\ref{profunc}). In order to visualize the result, all points
in space were plotted at which the modulus of the topological charge
density exceeds a certain value, namely $|q(x)| > 1$, for each time $t$.
The results are displayed in Figs.~\ref{fig7}-\ref{fig9}.

At the times $t=-0.3,\, 0,\, 0.3$, one finds a round lump of topological
charge density $q(x) > 1$ near the center of the configuration;
this is the contribution from the smeared-out intersection point.
Furthermore, there are peripheral kidney-shaped lumps with $q(x) < -1$.
Also the lumps seen at $t=-0.9,\, -0.6,\, 0.6,\, 0.9$ are associated
with $q(x) < -1$. These are all the topological charge
contributions induced by vortex writhe; they are evidently spread out 
smoothly over much of the vortex world-surface (and its surroundings),
corresponding to the smooth, gradual way in which the surface writhes.
Careful evaluation of the topological charge density using thickened
vortices thus allows one to recover all contributions to the topological
charge, consistent with the lattice analysis and general arguments.
Smoothness of a vortex surface does not preclude writhing contributions,
as may have been thought from naively considering infinitely thin surfaces
from the start, cf.~the discussion at the beginning of section
\ref{contsec}. Below, it will be shown that the writhing contributions
persist for infinitely thin vortex surfaces, by explicitly taking the
limit of a vanishing thickness, $a\rightarrow 0$.

\subsubsection{Thin vortex}
\noindent
Inspecting the expressions (\ref{qdens}) or (\ref{q3trig}) for the
topological charge density,
taking the thin vortex limit implies that contributions to $q(x)$ can only
come from points in the $s,s^{\prime } $-integrals where
$y(s)=y(s^{\prime } )$, since the profile functions become
$\delta $-distributions in this limit. In the case of the surface
parametrization (\ref{paramet}), the condition $y(s)=y(s^{\prime } )$
can be realized in two ways; either $s=s^{\prime } $, corresponding to
writhing contributions, or
$(s_2 , s_{2}^{\prime } ) \in \{ (0,\pi /2),(\pi /2,0) \}$ corresponding
to the intersection point at $y=0$
(note that $s_2 =s_{2}^{\prime} \in \{ 0,\pi /2 \} $ is already included
in the former case). Thus, in the space of parameters $s,s^{\prime } $,
the self-intersection contribution is conveniently isolated from the
writhe, $s=s^{\prime } $, and the corresponding regions in the integral
over $s,s^{\prime } $ can be treated separately.

Consider first the self-intersection point. In this case, as already
mentioned further above, one can work directly with thin vortices from the
start, i.e., one can straightforwardly substitute $f(z)=\delta^{4} (z)$ in
(\ref{q3trig}),
\begin{eqnarray}
q(x) &=& \frac{1}{4} \int d^2 s \int d^2 s^{\prime } \,
\delta^{4} (x-y(s) ) \delta^{4} (y(s)-y(s^{\prime } ) ) 
\ \cdot \label{q3thin} \\
& & \left[
\sin 2s_2 \sin 2s_{2}^{\prime } \sin^{2} (s_1 -s_{1}^{\prime } )
-\sin 3(s_2 -s_{2}^{\prime } ) \sin (s_2 -s_{2}^{\prime } ) \right]
\nonumber
\end{eqnarray}
where also $x$ has been substituted by $y(s)$ in the second $\delta $-function
in view of the presence of the first one. Now, consider in particular the
integration region where $s_2 $ is in the vicinity of $0$ and
$s_{2}^{\prime } $ is in the vicinity of $\pi /2$. In this limit, the
square brackets in (\ref{q3thin}) reduce to unity; furthermore, substituting
$s_2 = \rho^{2} /2 $ and $s_{2}^{\prime } = (\pi -\rho^{\prime \, 2} ) /2 $,
one has
\begin{equation}
y(s_1 , \rho^{2} /2 ) = \rho \left(
\begin{array}{c}
-\cos s_1 / \sqrt{2} \\ -\cos s_1 / \sqrt{2} \\ -\sin s_1 \\ 0
\end{array} \right)
\ \ \ \ \ \ \ \ \ \ \ \
y(s_{1}^{\prime } , (\pi -\rho^{\prime \, 2} ) /2 ) = \rho^{\prime } \left(
\begin{array}{c}
-\cos s_{1}^{\prime } / \sqrt{2} \\ \cos s_{1}^{\prime } / \sqrt{2} \\
0 \\ \sin s_{1}^{\prime }
\end{array} \right)
\end{equation}
respectively, to leading order in $\rho ,\rho^{\prime } $, where
$s_1 , s_{1}^{\prime } \in [0,2\pi ]$ as before. Therefore, (\ref{q3thin})
reduces to
\begin{eqnarray}
q(x) \! \! &=& \! \! \frac{1}{4} \int ds_1 \, ds_{1}^{\prime } \,
d\rho \, d\rho^{\prime } \,
\rho \, \rho^{\prime } \, \delta^{4} (x-y(s_1 ,\rho^{2} /2 ) ) \ \cdot \\
& & \delta \left( -(\rho \cos s_1 -\rho^{\prime } \cos s_{1}^{\prime } )/
\sqrt{2} \right)
\delta \left( -(\rho \cos s_1 +\rho^{\prime } \cos s_{1}^{\prime } )/
\sqrt{2} \right)
\delta \left( -\rho \sin s_1 \right) 
\delta \left( -\rho^{\prime } \sin s_{1}^{\prime } \right)
\nonumber
\end{eqnarray}
where the integrations over the (positive) variables $\rho ,\rho^{\prime } $
cover the vicinity of $\rho ,\rho^{\prime } =0$. Going to Cartesian
coordinates, $w_1 = \rho \cos s_1 $, $w_2 = \rho \sin s_1 $, and
correspondingly for the primed variables, one ends up with
\begin{eqnarray}
q(x) &=& \frac{1}{4} 
\int dw_1 \, dw_2 \, dw_{1}^{\prime } \, dw_{2}^{\prime } \,
\delta^{4} (x-y(s_1 (w_1 , w_2 ), s_2 (w_1 , w_2 ) ) ) \ \cdot 
\nonumber \\
& & \ \ \ \ \ \ \ \ \ \ \ \ \ \ \ \ \ \ \ \
\delta \left( -(w_1 - w_{1}^{\prime } )/ \sqrt{2} \right)
\delta \left( -(w_1 + w_{1}^{\prime } )/ \sqrt{2} \right)
\delta \left( -w_2 \right)
\delta \left( -w_{2}^{\prime } \right) \nonumber \\
&=& \frac{1}{4} \delta^{4} (x) \ .
\end{eqnarray}
Supplementing this with the integration region in (\ref{q3thin}) where the
roles of $s$ and $s^{\prime } $ are interchanged, i.e., the former is in
the vicinity of $\pi /2$ and the latter in the vicinity of $0$, one obtains
the same contribution again; thus, the topological charge density associated
with the vortex self-intersection point altogether is
\begin{equation}
q_{int} (x) = \frac{1}{2} \delta^{4} (x) \ .
\end{equation}
Correspondingly, by integrating over space-time, one obtains a contribution
$Q_{int} = 1/2$ to the global topological charge from the vortex
self-intersection point.
Note that, in the case of the intersection point, it was possible to work
with thin vortices from the start and the application of the formalism
was straightforward, yielding the correct result. In the case of the
writhing contributions, on the other hand, this is not the case; the
formalism becomes ambiguous if one naively uses thin vortices. To see
this, consider again eq.~(\ref{q3thin}). In the case $s=s^{\prime } $,
the term in the square brackets vanishes, corresponding to the fact that the
tangent space at any point of a smooth surface is 
two-dimensional\footnote{Note that, also in the general expression
(\ref{qdens}), the combination $\epsilon_{\rho \tau \beta \gamma }
\Sigma_{\rho \tau } (s) \Sigma_{\beta \gamma } (s^{\prime } )$ is
easily seen to vanish when $s=s^{\prime } $.}
(away from self-intersection points). On the other
hand, when $s=s^{\prime } $, the argument of the second $\delta $-function,
$\delta^{4} (y(s)-y(s^{\prime } ) )$, vanishes on a whole two-dimensional
sub-manifold of space-time. Two out of the four dimensions in this
$\delta $-function suffice to enforce $s=s^{\prime } $, while the remaining
two stay as $\delta^{2} (0)$. This is precisely the situation discussed in
the initial remarks of section \ref{contsec}; the product of these two
factors is ambiguous, and one has to work to higher order in the
thickness of the vortex in order to obtain a well-defined limit and thus
properly extract the writhing contributions. Starting again from
(\ref{q3trig}), one can write, in the limit $a\rightarrow 0$,
\begin{eqnarray}
q(x) &=& \frac{1}{4} \int d^2 s \, \delta^{4} (x-y(s) )
\ \cdot \label{q3writhe} \\
& & \int d^2 s^{\prime } \, f(y(s)-y(s^{\prime } ) ) \left[
\sin 2s_2 \sin 2s_{2}^{\prime } \sin^{2} (s_1 -s_{1}^{\prime } )
-\sin 3(s_2 -s_{2}^{\prime } ) \sin (s_2 -s_{2}^{\prime } ) \right] \ .
\nonumber
\end{eqnarray}
Note thus that $f(x-y(s) )$ in (\ref{q3trig}) has already been
substituted by its thin limit, $\delta^{4} (x-y(s) )$; this will be
justified a posteriori by the second line in (\ref{q3writhe}) yielding
a well-defined result. Correspondingly, in the remaining profile function $f$
in (\ref{q3writhe}), $x$ has again been substituted by $y(s)$ due to the
presence of $\delta^{4} (x-y(s) )$. Now, since the writhing contributions
originate from the region around $s=s^{\prime } $, it is useful to
substitute
\begin{equation}
s_{a}^{\prime } = s_a + r_a
\end{equation}
and to expand in the small quantities $r_a $; the profile function $f$
will restrict the new $r_a $-integrations to small $r_a $ when the vortex
thickness $a$ becomes small. To second order in the $r_a $, the square
bracket in (\ref{q3writhe}) reads
\begin{equation}
\sin 2s_2 \sin 2(s_2 +r_2 ) \sin^{2} r_1 -\sin 3r_2 \sin r_2
= r_1^2 \sin^{2} 2s_2 - 3r_2^2 + O(r^3 ) \ .
\end{equation}
Furthermore, the profile function $f$ depends on 
$(y(s)-y(s^{\prime } ) )^2 $, which to this order reads
\begin{equation}
(y(s)-y(s^{\prime } ) )^2 =
\left( \frac{\partial y(s)}{\partial s_a } r_a \right)^{2}
+ O(r^3 ) =
r_1^2 \sin 2s_2 + \frac{r_2^2 }{\sin 2s_2 } + O(r^3 )
\end{equation}
as can be verified straightforwardly by computing the scalar products of the
gradient vectors $\partial y(s)/\partial s_a $,
cf.~eqs.~(\ref{gradvec1}),(\ref{gradvec2}).
As a consequence, the topological charge density reduces to
\begin{eqnarray}
q(x) &=& \frac{1}{4} \int d^2 s \, \delta^{4} (x-y(s) )
\ \cdot \label{writhexp} \\
& & \frac{1}{a^4 \pi^{2} } \int dr_1 \, dr_2 \,
\exp \left( -\frac{1}{a^2 } 
\left( r_1^2 \sin 2s_2 + \frac{r_2^2 }{\sin 2s_2 } \right) \right)
\left( r_1^2 \sin^{2} 2s_2 - 3r_2^2 \right) \ .
\nonumber
\end{eqnarray}
In the limit $a\rightarrow 0$, the Gaussian strongly suppresses the integrand
for increasing $r_1 , r_2 $, so that the integration ranges of $r_1 , r_2 $
can be extended over the whole real axis without incurring an error (except
possibly for the boundary values $s_2 =0$ and $s_2 =\pi /2 $, at which the
integral will be defined by continuity and will presently be seen to vanish).
Furthermore, the higher order terms in the $r_a $ which have been neglected
above only lead to corrections which vanish as $a\rightarrow 0$. Evaluating
the Gaussian integrals in (\ref{writhexp}), one arrives at
\begin{equation}
q_{writhe} (x) = \int d^2 s \, \delta^{4} (x-y(s) )
\left( -\frac{1}{4\pi } \sin 2s_2 \right) \ .
\label{qwrithe}
\end{equation}
Thus, the topological charge density originating from vortex writhe persists
in the thin limit; it is concentrated on the vortex surface and, in view of
$(y(s))^2 = \sin 2s_2 $, grows as the square of the distance from the
origin, consistent with the visualization for thick vortices exhibited in
Figs.~\ref{fig7}-\ref{fig9}. The authors checked that the same result is
obtained using a (properly normalized) Lorentzian for the profile function
$f$. Furthermore, by integrating (\ref{qwrithe}) over space-time, one
obtains a writhing contribution to the global topological charge
\begin{equation}
Q_{writhe} = -\frac{1}{4\pi } \int_{0}^{2\pi } ds_1 \, 
\int_{0}^{\pi /2} ds_2 \, \sin 2s_2
= -\frac{1}{2} \ ,
\end{equation}
which exactly cancels
the contribution from the self-intersection point $Q_{int} $, as expected.

As a last point, it will be sketched how the above computation of the
writhing contribution can be generalized beyond the specific example
considered here, i.e., starting from the general expression (\ref{qdens})
instead of (\ref{q3trig}) and taking the limits $s^{\prime } \rightarrow s$,
$a\rightarrow 0$ in analogy to eqs.~(\ref{q3writhe})-(\ref{qwrithe}).
The general form analogous to (\ref{writhexp}) reads
\begin{eqnarray}
q(x) &=& \int d^2 s \, \delta^{4} (x-y(s) )
\ \cdot \label{general} \\
& & \frac{1}{16a^4 \pi^{2} } \int d^2 r \,
[0+O(r)+h^{ab}r_a r_b+O(r^3)] \exp\left(-g^{ab}r_a r_b/a^2 \right)
\nonumber
\end{eqnarray}
where the leading term is always absent due to the geometry of the expression
for the topological charge density, i.e., the combination
$\epsilon_{\rho \tau \beta \gamma } \Sigma_{\rho \tau } (s)
\Sigma_{\beta \gamma } (s)$, cf.~(\ref{qdens}), vanishes
identically\footnote{Note that the analogous term in
the Yang-Mills action diverges as $1/a^2 $, cf.~\cite{cont,kleinert}.}.
Here, the induced metric and a bilinear in gradients of the surface
elements have been introduced,
\begin{equation}
g^{ab} \equiv \frac{\partial y_\mu}{\partial s_a}
\frac{\partial y_\mu}{\partial s_b},\qquad
h^{ab} \equiv \Sigma_{\mu\nu}
\frac{\partial^2 \tilde{\Sigma}_{\mu\nu} }{\partial s_a \partial s_b}
=\frac{\partial \Sigma_{\mu\nu}}{\partial s_a}
\frac{\partial \tilde{\Sigma}_{\mu\nu} }{\partial s_b},\qquad
\tilde{\Sigma}_{\mu\nu}\equiv\frac{1}{2}
\epsilon_{\mu\nu\kappa\lambda}\Sigma_{\kappa\lambda}
\end{equation}
respectively. To obtain the second expression for $h^{ab} $, it has
been used that terms of the form
$\epsilon_{\mu\nu\kappa\lambda} (\partial y_{\mu }/\partial s_a )
(\partial y_{\nu }/\partial s_b ) (\partial y_{\kappa }/\partial s_c )$
vanish for all $\lambda, a, b, c$ since the derivatives contain either
$s_1$ or $s_2$ twice, rendering the expression symmetric in the
corresponding greek indices. The same argument is used again to 
arrive at (\ref{generalq}) below. As before, only the terms written
out explicitly in (\ref{general}) are relevant and evaluate to
\begin{equation}
q(x) = \int d^2 s \, \delta^{4} (x-y(s) ) \left(
-\frac{1}{16\pi } h^{ab}
\frac{\partial }{\partial g^{ab} }\frac{1}{\sqrt{g}} \right)
\end{equation}
where $g=\det g^{ab}$. This is the generalisation of (\ref{qwrithe}),
where, using the parametrization (\ref{paramet}), $g^{ab} $ and $h^{ab} $
happen to be diagonal and $g=1$. By virtue of the inverse metric $g_{ab}$,
one finally arrives at \cite{cont}
\begin{eqnarray}
q(x)&=& \frac{1}{32\pi } \int d^2 s\delta^4(x-y(s))
\left( \frac{1}{\sqrt{g}}g_{ab}
\frac{\partial \Sigma_{\mu\nu}}{\partial s_a}
\frac{\partial \tilde{\Sigma}_{\mu\nu}}{\partial s_b}\right)
\nonumber \\
&=&\frac{1}{32\pi }\int d^2 s\sqrt{g}\,\delta^4(x-y(s))g_{ab}\,
\frac{\partial}{\partial s_a}\frac{\Sigma_{\mu\nu}}{\sqrt{g}}
\frac{\partial}{\partial s_b}\frac{\tilde{\Sigma}_{\mu\nu}}{\sqrt{g}} \ ,
\label{generalq}
\end{eqnarray}
the latter being more natural from the reparametrization point
of view (where  $d^2 s\sqrt{g}$ and
$\Sigma/\sqrt{g}=\epsilon_{ab}/\sqrt{g}\ldots$ are the proper objects).
The space-time integral over this expression is known as a representation of
(minus) the self-intersection number, which is thus given in terms of an
integral over a smooth density \cite{polyakov}. In the context of the present
work, this identification corresponds to the observation that the ``smooth''
writhing contribution to the topological charge cancels the ``discrete'' one
from the self-intersection point. The validity of (\ref{generalq}) has been
corroborated in the present paper by considering the thick vortex explicitly.

There exists another intuitive formula for the self-intersection number
in terms of normal gauge fields \cite{mazur,pawel}, which is related to the
fact that $\tilde{\Sigma}$ is normal to the vortex surface: Take two
orthonormal normal vectors $n^1$ and $n^2$, and define an $SO(2)$ gauge
field $A=n^1_\mu {\rm d}n^2_\mu$ and its field strength $F={\rm d}A$.
Then the self-intersection number is given by $\int F/4\pi$.

As is best seen from the appearance of
$h\sim \partial \Sigma \partial \tilde{\Sigma } $ in the equations above,
the writhe is distributed according to the gradients of the tangent space and
of the normal space, respectively. The vortex example discussed in this work
is planar at the self-intersection point such that the writhing
contribution vanishes there, but this will not be the case for arbitrary
vortex surfaces.

\section{Concluding remarks}
\noindent
Using a specific simple example, it has been shown in this work how
topological charge arises through vortex writhe in the space-time
continuum. The topological charge density was visualized for a
thickened vortex and it was verified that the global topological charge is
independent of the vortex thickness. Topological charge through writhe
persists in the limit of arbitrarily thin vortex world-surfaces,
even though it is rather hidden in the singular nature of the vortex
field strength in this limit. Smoothness of a vortex surface does not
preclude these writhing contributions, which were demonstrated to be
proportional to a particular combination of gradients of the surface
elements\footnote{Note that the mere existence of such gradients does
not suffice; for instance, the 2-sphere embedded in $R^4 $ possesses
gradients of the surface elements, but the writhe vanishes.}, i.e., 
of the tangent space.

In the example discussed here, the appearance of writhe and the existence
of the self-intersection point are intertwined; a closed vortex surface
not extending to infinity has to bend and ``come back'' in order to display
a lone self-intersection point. This need not be the case when space-time
possesses compact directions; the vortex surface can then be closed due to
periodicity. Therefore, on the four-torus, one can have a (thin) vortex
consisting of two branches $x_1=x_2=0$ and $x_3=x_4=0$ which intersect
at a point but otherwise are completely planar. Then the contributions
to the topological charge are $Q_{int}=1/2$ and $Q_{writhe}=0$. This does
not contradict general statements on the quantization of topological
charge; half-integer topological charges exist on the torus when accompanied
by twisted boundary conditions\footnote{Indeed, the gauge fields
corresponding to this configuration,
$A_1=\pi\Theta(x_2)\delta(x_1)\sigma^3 $, 
$A_3=\pi\Theta(x_4)\delta(x_3)\sigma^3 $, $A_2=A_4=0$ obey transition
functions similar to the charge $1/2$ instanton solutions of constant
field strength \cite{thoowind,baalwind}.}. On the four-torus a single
self-intersection point can thus occur without writhe. Conversely,
writhe can obviously exist without a self-intersection point, a simple
example being the time evolution of Fig.~\ref{fig4} from $t=-0.9$
to $t=0-$ followed by the time reversed process (i.e., the loop is
unscrewed back in the direction it came from instead of being twisted
further as in Fig.~\ref{fig4}). Globally, this leads to $Q_{writhe}=0$.
An example of a (non-orientable) surface with writhing contribution
$Q_{writhe}=1/2$ and no intersection point is depicted in \cite{csb}.
A smooth continuum analogue of this surface (as well as the precise
relation to twisted boundary conditions) has not been given so far.

Two additional points became apparent in the course of this investigation
which deserve mention. For one, comparing sections \ref{latsec} and
\ref{contsec}, it is considerably simpler to evaluate the topological
charge of a surface made up of elementary squares on a hypercubic lattice
than to evaluate it for a general continuum surface, in which writhe is
rather hidden, as already remarked further above. On the lattice, topological
charge is concentrated at lattice sites as opposed to being smeared out
over the surface; as a consequence, it is simply accessible by counting
pairs of mutually orthogonal elementary squares meeting at the lattice
sites, cf.~also \cite{preptop}. A viable and efficient procedure of
determining the global topological charge of a general continuum surface
indeed would lie in latticizing it, i.e.~finding a topologically equivalent
lattice surface on a suitably fine lattice, and evaluating its global
topological charge instead\footnote{Note however that, as far as the
local topological density is concerned, the (absolute values of) local
contributions to the topological charge on rectangular lattices are bounded
from below by the ``quantum'' 1/16 (not 1/32, because at the sites of a
{\em closed} lattice surface, there cannot be a {\em single} pair of mutually
perpendicular elementary squares). Therefore, to approximate continuum
writhing contributions to the topological {\em density} to any precision,
one has to use non-rectangular polygonal surfaces, in the simplest case just
a triangulation of the continuum surface.}. Note that this stands in marked
contrast to the usual statement made for lattice Yang-Mills configurations;
on the standard Yang-Mills lattice, carrying Yang-Mills link variables,
topological charge is much harder to define and detect than in the
continuum, and indeed becomes an ambiguous quantity for generic Yang-Mills
configurations. By contrast, on a lattice carrying vortex fluxes such
as used here, which is dual to the standard Yang-Mills lattice, topological
charge actually is easier to evaluate than in the continuum.

Finally, note that a standard way of capturing the topology of lines in
three-dimensional space lies in defining a framing. One constructs a
second line displaced from the original one by a small distance; the
two lines define the edges of a ribbon which may twist and writhe, with
the two lines correspondingly entangled. With the help of such a framing,
one can define a (three-dimensional) writhing number; the topological
charge associated with the whole world-surface of a vortex line can then
be related to the difference between the writhing numbers at the initial
and final times bounding the space-time region under consideration
\cite{cont}. The procedure of defining a framing fits into the formalism
used in the present work; it corresponds to one particular way of choosing
the profile function $f$. Namely, the original thin vortex line at a given
time, described by a $\delta $-function profile, is ``smeared'' into two
thin lines, i.e.~two $\delta $-functions displaced from one another by a
small distance. For the purposes of the present investigation, it was
however more appropriate to choose a smoother profile function, leading to
a finite gluonic field strength; not only from a technical point of view,
but also from a physics point of view, this is a natural choice, since
realistic physical vortices indeed possess thick, spread-out transverse
profiles.

\section*{Acknowledgments}
M.E.~acknowledges the hospitality of P.~van Baal and the Instituut-Lorentz
at Leiden University, where this work was initiated.
R.~Bertle and M.~Faber are acknowledged for stimulating
discussions which yielded the questions which the present work aims to
answer, as well as for providing the MATHEMATICA routine with which
Figs.~\ref{fig1} and \ref{fig6} were generated. F.B.~thanks Ph.~de~Forcrand,
M.~Pepe, A.~Wipf and, in particular, O.~Jahn for discussions on the
non-Abelian version of the vortex discussed here. Finally, the authors are
also grateful to science+computing ag, T\"ubingen, for providing
computational resources.

\begin{appendix}
\section{Vortex loop at fixed time}
\label{apploop}
\noindent
In order to cast the parametrization (\ref{paramet}),
\[
y (s_1 , s_2 ) = \sqrt{\sin 2s_2 } \left(
\begin{array}{c}
-\cos s_1 \ (\cos s_2 + \sin s_2 ) / \sqrt{2} \\
-\cos s_1 \ (\cos s_2 - \sin s_2 ) / \sqrt{2} \\
-\sin s_1 \ \cos s_2 \\
\sin s_1 \ \sin s_2
\end{array} \right)
\]
in terms of the time evolution of a closed loop in three-dimensional space,
as plotted in Fig.~\ref{fig4} for selected times $t$, one eliminates the 
parameter $s_1 $ in favor of the time
\begin{equation}
t\equiv y_1 (s_1 , s_2 ) =
-\sqrt{\sin 2s_2 } \cos s_1 \ (\cos s_2 + \sin s_2 ) / \sqrt{2}
\ \ \in \ [-1,1]
\end{equation}
i.e.,
\begin{eqnarray}
\cos s_1 &=&
-\frac{\sqrt{2} \, t}{\sqrt{\sin 2s_2 } (\cos s_2 + \sin s_2 )} \\
\sin s_1 &=& \pm \sqrt{1-\cos^{2} s_1 }
\end{eqnarray}
where both signs are relevant; the plus sign yields the solution found in
the interval $s_1 \in [0,\pi ]$, whereas the minus sign yields the solution
found in the interval $s_1 \in [\pi , 2\pi ]$. Both solutions must be taken
into account in order to reproduce the parametrization (\ref{paramet}) in
the entire interval $s_1 \in [0,2\pi ]$. Thus, one obtains the alternative
parametrization
\begin{equation}
y (t, s_2 ) = \left( \begin{array}{c} t \\
t(\cos s_2 - \sin s_2 )/(\cos s_2 + \sin s_2 ) \\
\mp \cos s_2 \sqrt{\sin 2s_2 - 2t^2 /(\cos s_2 + \sin s_2 )^2 } \\
\pm \sin s_2 \sqrt{\sin 2s_2 - 2t^2 /(\cos s_2 + \sin s_2 )^2 }
\end{array} \right)
\label{paralt}
\end{equation}
which, at fixed time $t$, parametrizes a closed loop via the parameter
$s_2 $, taking into account both choices of sign. Note that, for a given
time $t$, the range of $s_2 $ is not anymore $s_2 \in [0, \pi /2 ]$, as in
the original parametrization (\ref{paramet}), but it is in general restricted
to a smaller interval by the requirement that the argument of the square
roots in (\ref{paralt}) be positive, i.e.,
\begin{equation}
2t^2 \le \sin 2s_2 (\cos s_2 + \sin s_2 )^2 = \sin 2s_2 (1+\sin 2s_2 ) \ .
\label{poscond}
\end{equation}
The condition thus reduces to a quadratic equation for $\sin 2s_2 $;
picking out the relevant solution by taking into account the requirement
$| \sin 2s_2 | \le 1$, eq.~(\ref{poscond}) is solved by
\begin{equation}
\sin 2s_2 \ge \frac{1}{2} (\sqrt{1+8t^2 } -1) \ .
\end{equation}
Thus,
\begin{equation}
s_2 \in \left[ \frac{1}{2} \arcsin \frac{1}{2} (\sqrt{1+8t^2 } -1) ,
\frac{\pi }{2} - \frac{1}{2} \arcsin \frac{1}{2} (\sqrt{1+8t^2 } -1) \right]
\end{equation}
at the time $t$.

\section{Continuum vortex from Laplacian Center Gauge instanton}
\label{pertinst}
\noindent
Abelian \cite{thooabel} and center \cite{mcg} gauges are used to define
magnetic monopoles and center vortices as defects of gauge fixings. In the
particular cases of the Laplacian Abelian Gauge \cite{vandersijs} and the
Laplacian Center Gauge \cite{forclap}, monopoles and vortices are defined,
respectively, to be the nodes of the ground state of the covariant
Laplacian $-\D^2[A]$ (in the adjoint representation) and the set of points
where the two lowest-lying eigenvectors of this operator are parallel. The
ground state $\phi$ of $-\D^2[A]$ in the background of a single instanton (in
regular gauge) has been found \cite{monhopf} to be threefold degenerate and
of the form ($1_2 $ denoting the $2\times 2$ unit matrix)
\begin{eqnarray}
\phi_a=r^2\,g^\dagger \sigma_a g,
\qquad g=(x_4 1_2+ix_a\sigma_a)/r
\end{eqnarray}
near the four-dimensional origin. There, both the monopole and the vortex
are located, degenerate to a pointlike defect.

In \cite{hopfseeds}, it was shown that perturbing the instanton background
$A$ such that the ground state is perturbed as $\phi_3\to\phi_3-R^2\sigma_3$
gives rise to a monopole loop of radius $R$ in the $x_1x_2$-plane.

For the center vortex, one considers in the same spirit the perturbation
\begin{eqnarray}
\phi_1^{\rm pert}=\phi_1+\sigma_3,\qquad\phi_3^{\rm pert}=\phi_3
\end{eqnarray}
where the relevant scale has been put to 1.
Straightforward algebra shows that $\phi_1^{\rm pert}$ and $\phi_3^{\rm pert}$
are parallel at
\begin{eqnarray}
\varphi_{12}-\varphi_{34}=\pi\:{\rm mod}\:2\pi,\qquad r=\sqrt{\sin2\theta}
\label{requirement}
\end{eqnarray}
where double polar coordinates have been introduced,
\begin{equation}
x_{\mu } =r(\cos\theta\cos\varphi_{12},
\cos\theta\sin\varphi_{12},
\sin\theta\cos\varphi_{34},
\sin\theta\sin\varphi_{34}) \ .
\end{equation}
This leaves
\begin{eqnarray}
s_2\equiv\theta\in[0,\pi/2],\qquad 
s_1\equiv(\varphi_{12}+\varphi_{34})/2\in[0,2\pi] 
\end{eqnarray}
as free parameters. Up to rotations and inversions in space-time, this
corresponds to the vortex surface parametrization of eq.~(\ref{paramet});
namely, one must permute the 1-, 3- and 4-components as 
$(1,3,4)\rightarrow (4,1,3)$, and then rotate the resulting 1- and 2-components
as $(x_1 ,x_2 ) \rightarrow ( (x_1 -x_2 )/\sqrt{2} , (-x_1 -x_2 )/\sqrt{2} )$.
Note that, without the latter rotation, the parametrization would have a
more symmetric form; however, for the purposes of visualization, the
choice of coordinates (\ref{paramet}) is more advantageous. This simply
amounts to a change in the point of view of the observer, not an actual
change in the form of the vortex surface.

The monopole defined by $\phi_1^{\rm pert}$ has to fulfil (\ref{requirement})
plus the condition $\theta=\pi/4$, i.e. it is a loop on the vortex which is
as far away as possible from the origin. The monopole of $\phi_3^{\rm pert}$
is still degenerate to the origin (but adding a perturbation to it lifts also
this degeneracy while leading to similar vortex surfaces). For the present
purposes, the detailed form of the monopole loops is not relevant; the
monopole content of the configuration resulting from the above construction
is not adopted in the body of this work. Only the surface (\ref{paramet}),
which in itself is orientable and thus is not forced to carry monopoles, is
used, and endowed with an oriented, smooth field strength.

\section{Bianchi identity}
\label{appbian}
\noindent
The field strength (\ref{fsdef}) satisfies the Abelian Bianchi identity
(\ref{bianchi}) independent of the profile function $f$.
Inserting (\ref{fsdef}) into (\ref{bianchi}) and using
$\epsilon_{\rho \tau \mu \nu } \epsilon_{\mu \nu \kappa \lambda }
=2(\delta_{\rho \kappa } \delta_{\tau \lambda }
- \delta_{\rho \lambda } \delta_{\tau \kappa } )$, one arrives at
\begin{equation}
\epsilon_{\rho \tau \mu \nu } \partial_{\tau } F_{\mu \nu } =
\pi \sigma^{3} \partial_{\tau } \int d^2 s \, f(x-y(s_1 , s_2 ) )
\left( \epsilon_{ab} \frac{\partial y_{\rho } }{\partial s_a }
\frac{\partial y_{\tau } }{\partial s_b }
- \epsilon_{ab} \frac{\partial y_{\tau } }{\partial s_a }
\frac{\partial y_{\rho } }{\partial s_b } \right) \ .
\end{equation}
Exchanging the dummy indices $a,b$ in the second term in the 
parenthesis, using $\epsilon_{ba} = -\epsilon_{ab} $, and furthermore
inserting
\begin{equation}
\frac{\partial y_{\tau } }{\partial s_b } \partial_{\tau } f(x-y(s_1 ,s_2 ))
= -\frac{\partial y_{\tau } }{\partial s_b } 
\frac{\partial }{\partial y_{\tau } } f(x-y(s_1 ,s_2 ))
= -\frac{\partial }{\partial s_b } f(x-y(s_1 ,s_2 ))
\end{equation}
yields
\begin{equation}
\epsilon_{\rho \tau \mu \nu } \partial_{\tau } F_{\mu \nu } = -2 \pi
\sigma^{3} \int ds_1 \, ds_2 \, \frac{\partial y_{\rho } }{\partial s_1 }
\frac{\partial }{\partial s_2 } f(x-y) +2 \pi \sigma^{3}
\int ds_1 \, ds_2 \, \frac{\partial y_{\rho } }{\partial s_2 }
\frac{\partial }{\partial s_1 } f(x-y)
\label{partarg}
\end{equation}
after having written out the sums over $a,b$ explicitly. Now, one can
partially integrate the first term over $s_2 $ and the second term over
$s_1 $ without generating surface contributions, upon which the two terms
are seen to cancel, as was to be shown. When partially integrating over
$s_1 $, there is no surface contribution because $y(0,s_2 ) = y(2\pi ,s_2 )$
for any $s_2 $. When partially integrating over $s_2 $, both
$y(s_1 ,0) = 0$ and $y(s_1 ,\pi /2) = 0$; therefore, also
$\partial y_{\rho } / \partial s_1 = 0$ for $s_2 = 0,\pi /2 $, i.e. the
surface contribution already vanishes separately for these two values
of $s_2 $.

Note that this argument does not particularly depend on the specifics of
the vortex surface considered here. Quite generally, since vortex
world-surfaces are closed, they can be viewed as ``time'' evolutions of
closed loops. Therefore, one can choose\footnote{This ``time'' does not
have to correspond to physical time; also in the case of the parametrization
(\ref{paramet}), the parameter $s_2 $ is not identical to physical time.}
``time'' as one parameter of the vortex surface, and an angle parametrizing
the closed loop at a given ``time'' as the other. As a result, partial
integration over the aforementioned angle never generates a surface
contribution, since the two end points of the interval on which the angle
is defined map to identical space-time points for any given ``time''.
On the other hand, at the initial and final ``times'', the surface
parametrization reduces to a point; otherwise, the surface would not be
closed. In other words, at these two ``times'', the surface parametrization
is a constant space-time point independent of the angle variable. As a
result, the derivative of the parametrization with respect to the angle,
which appears in the surface pieces under consideration (cf.~the first
term in (\ref{partarg})), vanishes.

\section{Center phase induced by vortex flux}
\label{appcent}
\noindent
To see that the normalization of (\ref{fsdef}) is adequately chosen such
that vortices indeed carry flux corresponding to the center of the $SU(2)$
gauge group, consider for the sake of the argument a planar vortex
world-surface extending into the 1-2-directions in space-time,
\begin{equation}
\bar{y} (s_1 , s_2 ) = (s_1 , s_2 , 0 , 0) \ , \ \ \ \ \
s_1 , s_2 \in [-\infty , \infty ] \ .
\label{planvor}
\end{equation}
Any smooth vortex surface can be viewed as being locally planar, i.e. of
the form (\ref{planvor}) on sufficiently small length scales (if the
space-time axes are chosen suitably). Inserting (\ref{planvor}) into
(\ref{fsdef}), one obtains the field strength
\begin{equation}
F_{34} (x) = \pi \sigma^{3}  \int d^2 s \, f(x-\bar{y} (s_1 , s_2 ) )
=\frac{\sigma^{3} }{a^2 } e^{-(x_3^2 + x_4^2 )/a^2 }
\end{equation}
where the explicit form (\ref{profunc}) has been inserted. Now, consider
a Wilson loop $C$ located in the 3-4-plane circumscribing the vortex
(e.g. a circle centered at the origin with sufficiently large radius $R$
to capture the entire vortex flux, which may be smeared out via the profile
function $f$; i.e., $R^2 /a^2 \gg 1 $),
\begin{equation}
W [C] = \frac{1}{2} \mbox{Tr} \,
\exp \left( i \oint_{C} dx_{\mu } A_{\mu } \right) =
\frac{1}{2} \mbox{Tr} \,
\exp \left( i \int_{S} dx_3 \, dx_4 \, F_{34} \right) =
\frac{1}{2} \mbox{Tr} \, e^{i\pi \sigma^{3} } = -1 \ .
\end{equation}
Here, for the second equality, Stokes' theorem has been used (in a fixed
time slice, $x_1 = 0$), i.e. $S$ represents the area in the 3-4-plane
bounded by $C$. For the third equality, the condition $R^2 /a^2 \gg 1 $ has
been used, which permits extending the integral over $S$ to the entire
3-4-plane without appreciably changing the result. Note also that no path
ordering is necessary in the Wilson loop due to the Abelian nature of the
problem. Thus, one indeed has $W [C] = -1$, i.e. the total flux carried by
the vortex corresponds to the nontrivial center element of the $SU(2)$
gauge group. Continuity of flux, i.e. the Bianchi identity (\ref{bianchi}),
guarantees that this result extends to the entirety of any vortex
world-surface given that it is true locally for a small vortex segment,
as shown explicitly here.

\section{Simplifying the topological charge density}
\label{appess}
\noindent
The topological charge density (\ref{qdens}),
\[
q(x) = \frac{1}{16} \int d^2 s \int d^2 s^{\prime } \,
\epsilon_{\rho \tau \beta \gamma } \,
\Sigma_{\rho \tau } (s) \Sigma_{\beta \gamma } (s^{\prime } )
f(x-y(s) ) f(x-y(s^{\prime } ) )
\]
where, cf.~(\ref{sufel}),
\[
\Sigma_{\kappa \lambda } (s) = \epsilon_{ab}
\frac{\partial y_{\kappa } }{\partial s_a }
\frac{\partial y_{\lambda } }{\partial s_b }
\]
can be rewritten using
\begin{equation}
\epsilon_{\rho \tau \beta \gamma } \, \epsilon_{ab} \,
\frac{\partial y_{\rho } (s) }{\partial s_a }
\frac{\partial y_{\tau } (s) }{\partial s_b } =
2\epsilon_{\rho \tau \beta \gamma } \,
\frac{\partial y_{\rho } (s) }{\partial s_1 }
\frac{\partial y_{\tau } (s) }{\partial s_2 }
\label{contracts}
\end{equation}
(and analogously for the sum over $c,d$ in conjunction with the derivatives
with respect to $s_{c}^{\prime } , s_{d}^{\prime } $) with the result
\begin{eqnarray}
q(x) &=& \frac{1}{2} \int d^2 s \int d^2 s^{\prime } \,
f(x-y(s) ) f(x-y(s^{\prime } ) ) \ \cdot 
\label{qexpli} \\
& & \ \ \ \left[
\left( \frac{\partial y_1 (s) }{\partial s_1 }
\frac{\partial y_2 (s) }{\partial s_2 } -
\frac{\partial y_2 (s) }{\partial s_1 }
\frac{\partial y_1 (s) }{\partial s_2 } \right)
\left( \frac{\partial y_3 (s^{\prime } ) }{\partial s_{1}^{\prime } }
\frac{\partial y_4 (s^{\prime } ) }{\partial s_{2}^{\prime } } -
\frac{\partial y_4 (s^{\prime } ) }{\partial s_{1}^{\prime } }
\frac{\partial y_3 (s^{\prime } ) }{\partial s_{2}^{\prime } } \right)
\right. \nonumber \\
& & \ \ \ \ +
\left( \frac{\partial y_1 (s) }{\partial s_1 }
\frac{\partial y_3 (s) }{\partial s_2 } -
\frac{\partial y_3 (s) }{\partial s_1 }
\frac{\partial y_1 (s) }{\partial s_2 } \right)
\left( \frac{\partial y_4 (s^{\prime } ) }{\partial s_{1}^{\prime } }
\frac{\partial y_2 (s^{\prime } ) }{\partial s_{2}^{\prime } } -
\frac{\partial y_2 (s^{\prime } ) }{\partial s_{1}^{\prime } }
\frac{\partial y_4 (s^{\prime } ) }{\partial s_{2}^{\prime } } \right) 
\nonumber \\
& & \ \ \ \ + \left.
\left( \frac{\partial y_2 (s) }{\partial s_1 }
\frac{\partial y_3 (s) }{\partial s_2 } -
\frac{\partial y_3 (s) }{\partial s_1 }
\frac{\partial y_2 (s) }{\partial s_2 } \right)
\left( \frac{\partial y_1 (s^{\prime } ) }{\partial s_{1}^{\prime } }
\frac{\partial y_4 (s^{\prime } ) }{\partial s_{2}^{\prime } } -
\frac{\partial y_4 (s^{\prime } ) }{\partial s_{1}^{\prime } }
\frac{\partial y_1 (s^{\prime } ) }{\partial s_{2}^{\prime } } \right)
\right] \nonumber
\end{eqnarray}
where the sums over $\rho , \tau , \beta , \gamma $ have been written out
explicitly and the number of terms has been halved by using the freedom
in exchanging the names of the dummy integration variables $s,s^{\prime } $.
Inserting the derivatives
\begin{equation}
\frac{\partial y (s_1 , s_2 )}{\partial s_1 }
= \sqrt{\sin 2s_2 } \left( \begin{array}{c}
\sin s_1 \ (\cos s_2 + \sin s_2 ) / \sqrt{2} \\
\sin s_1 \ (\cos s_2 - \sin s_2 ) / \sqrt{2} \\
-\cos s_1 \ \cos s_2 \\
\cos s_1 \ \sin s_2
\end{array} \right)
\label{gradvec1}
\end{equation}
and
\begin{equation}
\frac{\partial y (s_1 , s_2 )}{\partial s_2 }
= \frac{1}{\sqrt{\sin 2s_2 } } \left( \begin{array}{c}
-\cos s_1 \ (\cos s_2 - \sin s_2 ) (1+4\cos s_2 \sin s_2 ) / \sqrt{2} \\
-\cos s_1 \ (\cos s_2 + \sin s_2 ) (1-4\cos s_2 \sin s_2 ) / \sqrt{2} \\
-\sin s_1 \ \cos s_2 \ (1-4\sin^{2} s_2 ) \\
-\sin s_1 \ \sin s_2 \ (1-4\cos^{2} s_2 )
\end{array} \right)
\label{gradvec2}
\end{equation}
(where $\cos 2s_2 $, $\sin 2s_2 $ have been expanded in terms of
$\cos s_2 $, $\sin s_2 $) into the six factors appearing in the square
brackets in (\ref{qexpli}), one has
\begin{eqnarray}
\frac{\partial y_1 (s) }{\partial s_1 }
\frac{\partial y_2 (s) }{\partial s_2 } -
\frac{\partial y_2 (s) }{\partial s_1 }
\frac{\partial y_1 (s) }{\partial s_2 } &=&
2\sin s_1 \ \cos s_1 \ \sin s_2 \ \cos s_2 \\
\frac{\partial y_3 (s^{\prime } ) }{\partial s_{1}^{\prime } }
\frac{\partial y_4 (s^{\prime } ) }{\partial s_{2}^{\prime } } -
\frac{\partial y_4 (s^{\prime } ) }{\partial s_{1}^{\prime } }
\frac{\partial y_3 (s^{\prime } ) }{\partial s_{2}^{\prime } } &=&
-2\sin s_{1}^{\prime } \ \cos s_{1}^{\prime } \ 
\sin s_{2}^{\prime } \ \cos s_{2}^{\prime } \\
\frac{\partial y_1 (s) }{\partial s_1 }
\frac{\partial y_3 (s) }{\partial s_2 } -
\frac{\partial y_3 (s) }{\partial s_1 }
\frac{\partial y_1 (s) }{\partial s_2 } &=&
-\frac{1}{\sqrt{2} } \cos s_2 \ \left(
\cos s_2 \ (1-4\sin^{2} s_2 ) \right. \label{combo1} \\
& & \ \ \ \ \ \ \ \ \ \ \ \ \ \ \ \ \ \ \left.
+ \sin s_2 \ (1-4\sin^{2} s_2 + 2\cos^{2} s_1 ) \right) \nonumber \\
\frac{\partial y_4 (s^{\prime } ) }{\partial s_{1}^{\prime } }
\frac{\partial y_2 (s^{\prime } ) }{\partial s_{2}^{\prime } } -
\frac{\partial y_2 (s^{\prime } ) }{\partial s_{1}^{\prime } }
\frac{\partial y_4 (s^{\prime } ) }{\partial s_{2}^{\prime } } &=&
\frac{1}{\sqrt{2} } \sin s_{2}^{\prime } \ \left(
-\sin s_{2}^{\prime } \ (1-4\cos^{2} s_{2}^{\prime } ) \right. 
\label{combo2} \\
& & \ \ \ \ \ \ \ \ \ \ \ \ \ \ \ \ \ \ \left.
+ \cos s_{2}^{\prime } \ (1-4\cos^{2} s_{2}^{\prime }
+ 2\cos^{2} s_{1}^{\prime } ) \right) \nonumber \\
\frac{\partial y_2 (s) }{\partial s_1 }
\frac{\partial y_3 (s) }{\partial s_2 } -
\frac{\partial y_3 (s) }{\partial s_1 }
\frac{\partial y_2 (s) }{\partial s_2 } &=&
-\frac{1}{\sqrt{2} } \cos s_2 \ \left(
\cos s_2 \ (1-4\sin^{2} s_2 ) \right. \label{combo3} \\
& & \ \ \ \ \ \ \ \ \ \ \ \ \ \ \ \ \ \ \left.
- \sin s_2 \ (1-4\sin^{2} s_2 + 2\cos^{2} s_1 ) \right) \nonumber \\
\frac{\partial y_1 (s^{\prime } ) }{\partial s_{1}^{\prime } }
\frac{\partial y_4 (s^{\prime } ) }{\partial s_{2}^{\prime } } -
\frac{\partial y_4 (s^{\prime } ) }{\partial s_{1}^{\prime } }
\frac{\partial y_1 (s^{\prime } ) }{\partial s_{2}^{\prime } } &=&
\frac{1}{\sqrt{2} } \sin s_{2}^{\prime } \ \left(
-\sin s_{2}^{\prime } \ (1-4\cos^{2} s_{2}^{\prime } ) \right. 
\label{combo4} \\
& & \ \ \ \ \ \ \ \ \ \ \ \ \ \ \ \ \ \ \left.
- \cos s_{2}^{\prime } \ (1-4\cos^{2} s_{2}^{\prime }
+ 2\cos^{2} s_{1}^{\prime } ) \right) \ . \nonumber
\end{eqnarray}
Comparing the right hand sides of eqs.~(\ref{combo1}) and (\ref{combo3}),
one notices that they can be cast in the forms $(A+B)$ and $(A-B)$
respectively, with expressions $A,B$ common to the two equations. Likewise,
the right hand sides of (\ref{combo2}) and (\ref{combo4}) can be cast in the
forms $(C+D)$ and $(C-D)$ respectively, with common expressions $C,D$. As a
result, after inserting (\ref{combo1})-(\ref{combo4}), the second and the
third lines in the square brackets in (\ref{qexpli}) correspond to the
forms $(A+B)(C+D)$ and $(A-B)(C-D)$ respectively, and therefore their sum
simplifies to $(A+B)(C+D) + (A-B)(C-D) = 2AC+2BD$. Thus, one has
\begin{eqnarray}
q(x) &=& \frac{1}{2} \int d^2 s \int d^2 s^{\prime } \,
f(x-y(s) ) f(x-y(s^{\prime } ) ) \ \cdot \label{qtrig} \\
& & \ \ \ \left[
-\frac{1}{4} \sin 2s_1 \sin 2s_2 \sin 2s_{1}^{\prime } \sin 2s_{2}^{\prime }
+\cos^{2} s_2 \sin^{2} s_{2}^{\prime }
(1-4\sin^{2} s_2 ) (1-4\cos^{2} s_{2}^{\prime } ) \right. \nonumber \\
& & \ \ \ \ \, \left.
-\frac{1}{4} \sin 2s_2 \sin 2s_{2}^{\prime }
(1-4\sin^{2} s_2 + 2\cos^{2} s_1 )
(1-4\cos^{2} s_{2}^{\prime } + 2\cos^{2} s_{1}^{\prime } )
\right] \nonumber
\end{eqnarray}
where the second term in the second line corresponds to the combination
$2AC$ referred to above, whereas the third line corresponds to $2BD$.
Of course, the first term in the second line simply comes from the first
line in the square brackets in (\ref{qexpli}). To further simplify
(\ref{qtrig}), it is useful to substitute
\begin{eqnarray}
\cos^{2} z &=& (1+\cos 2z)/2 \label{cosdoub} \\
\sin^{2} z &=& (1-\cos 2z)/2
\end{eqnarray}
everywhere (with $z$ denoting any of the variables $s_1 , s_{1}^{\prime } ,
s_2 , s_{2}^{\prime } $). Multiplying out the resulting terms in the
square brackets in (\ref{qtrig}), one finds
\begin{eqnarray}
q(x)\! &=& \! \frac{1}{8} \int d^2 s \int d^2 s^{\prime } \,
f(x-y(s) ) f(x-y(s^{\prime } ) ) \ \cdot \label{q2trig} \\
& & \left[
-\sin 2s_2 \sin 2s_{2}^{\prime } \sin 2s_1 \sin 2s_{1}^{\prime }
\right. \nonumber \\
& &
+1-(\cos 2s_2 -\cos 2s_{2}^{\prime } ) - \cos 2s_2 \cos 2s_{2}^{\prime }
-2\cos^{2} 2s_2 - 2\cos^{2} 2s_{2}^{\prime } \nonumber \\
& & \mbox{\hspace{4.7cm} }
-2\cos 2s_2 \cos 2s_{2}^{\prime } (\cos 2s_2 - \cos 2s_{2}^{\prime } )
+4\cos^{2} 2s_2 \cos^{2} 2s_{2}^{\prime } \nonumber \\
& &
-\sin 2s_2 \sin 2s_{2}^{\prime } \cos 2s_1 \cos 2s_{1}^{\prime }
-2\sin 2s_2 \sin 2s_{2}^{\prime }
(\cos 2s_2 \cos 2s_{1}^{\prime } - \cos 2s_1 \cos 2s_{2}^{\prime } )
\nonumber \\
& & \left. \mbox{\hspace{8.7cm} }
+4\sin 2s_2 \sin 2s_{2}^{\prime } \cos 2s_2 \cos 2s_{2}^{\prime }
\right] \nonumber
\end{eqnarray}
where a factor four has been extracted from the square brackets.
Now, all the terms inside the square brackets which are grouped into pairs
by parentheses are seen to cancel if one uses the freedom to exchange the
names of the dummy integration variables $s,s^{\prime } $ on one of the
members of each pair. Furthermore, by using
\begin{equation}
\cos 2s_1 \cos 2s_{1}^{\prime } + \sin 2s_1 \sin 2s_{1}^{\prime }
= \cos 2(s_1 - s_{1}^{\prime } )
= 1-2\sin^{2} (s_1 - s_{1}^{\prime } )
\end{equation}
to combine the first terms of the first and the fourth lines in the square
brackets in (\ref{q2trig}), and supplementing this with the contribution
$-\cos 2s_2 \cos 2s_{2}^{\prime } $ from the second line, one obtains
\begin{eqnarray}
q(x) &=& \frac{1}{8} \int d^2 s \int d^2 s^{\prime } \,
f(x-y(s) ) f(x-y(s^{\prime } ) ) \ \cdot \label{q4trig} \\
& & \left[
2\sin 2s_2 \sin 2s_{2}^{\prime } \sin^{2} (s_1 -s_{1}^{\prime } )
-\cos 2 (s_2 -s_{2}^{\prime } ) + \sin 4s_2 \sin 4s_{2}^{\prime }
\right. \nonumber \\
& & \left.
+(1-2\cos^{2} 2s_2 ) \, (1-2\cos^{2} 2s_{2}^{\prime } ) \right]
\nonumber
\end{eqnarray}
after having combined the remaining terms in the second and third lines
in the square brackets in (\ref{q2trig}) into the product in the last line
of (\ref{q4trig}). Again using (\ref{cosdoub}), this time with
$z=2s_2 , 2s_{2}^{\prime } $, one finally arrives at the result aimed for,
eq.~(\ref{q3trig}),
\begin{eqnarray*}
q(x) &=& \frac{1}{4} \int d^2 s \int d^2 s^{\prime } \,
f(x-y(s) ) f(x-y(s^{\prime } ) ) \ \cdot \\
& & \left[
\sin 2s_2 \sin 2s_{2}^{\prime } \sin^{2} (s_1 -s_{1}^{\prime } )
-\sin 3(s_2 -s_{2}^{\prime } ) \sin (s_2 -s_{2}^{\prime } ) \right]
\nonumber
\end{eqnarray*}
after having used
\begin{equation}
\cos 4 (s_2 -s_{2}^{\prime } ) -\cos 2 (s_2 -s_{2}^{\prime } )
= -2 \sin 3(s_2 -s_{2}^{\prime } ) \sin (s_2 -s_{2}^{\prime } )
\end{equation}
and having extracted a factor two from the square brackets.

\end{appendix}

\end{document}